\documentclass[aps,11pt,prd]{revtex4}
\usepackage{graphicx}
\usepackage{cancel}
\usepackage{amssymb}
\usepackage{textcomp}
\usepackage{amsmath}
\usepackage{bm}
\usepackage{times}
\usepackage{epsfig}
\usepackage{color}
\usepackage{graphics}
\usepackage{hyperref}
\usepackage{setspace}
\usepackage{paralist}
\usepackage{placeins}
\hypersetup{
    pdfnewwindow=true,      
    colorlinks=true,        
    linkcolor=black,        
    citecolor=blue,         
    filecolor=blue,         
    urlcolor=blue           
}

\usepackage{epsfig}
\usepackage{color}
\usepackage{slashed}
\usepackage{comment}
\usepackage{epstopdf}
\usepackage{array}
\usepackage{tikz}

\begin{document}

\title{\Large {\bf{Lepton Flavour Violation in Left-Right Theory}}}
\author{Pavel Fileviez P\'erez$^{1}$, Clara Murgui$^{2}$}
\affiliation{$^{1}$CERCA, Physics Department, Case Western Reserve University,
Rockefeller Bldg. 2076 Adelbert Rd. Cleveland, OH 44106, USA \\
$^{2}$Departamento de F\'isica Te\'orica, IFIC, Universitat de Valencia-CSIC, 
E-46071, Valencia, Spain}
%
\begin{abstract}
We investigate the predictions for lepton flavour number violating processes in the context of a simple left-right symmetric theory. 
In this context neutrinos are Majorana fermions and their masses are generated at the quantum level through the Zee mechanism using the simplest Higgs sector.
We show that the right handed neutrinos are generically light and can give rise to large lepton flavour violating contributions to rare processes. 
We discuss the correlation between the collider constraints and the predictions for lepton flavour violating processes. We find that using the predictions 
for $\mu \to e \gamma$ and $\mu \to e$ conversion together with the collider signatures one could test this theory in the near future. 
\end{abstract}
%
\maketitle
%
%

\section{Introduction}
The Large Hadron Collider (LHC) has discovered the last missing piece of the Standard Model (SM) of particle physics.
The discovery of the Brout-Englert-Higgs boson was crucial to establish the SM as one of the most important theories of nature. 
Today, we believe that the SM should be an effective theory to explain most of the current experimental results. However, it is well-known 
that one cannot explain in this context, for example, the hierarchy of the fermion masses, the origin of neutrino masses, the origin of Parity and CP violation, the nature of dark matter and 
the baryon-asymmetry in the Universe. 

There are many ideas for physics beyond the Standard Model which can help us to define a new theory to describe the new energy scale, TeV scale, 
which is currently explored by the Large Hadron Collider. In the context of left-right symmetric theories~\cite{Pati:1974yy,Mohapatra:1974gc,Senjanovic:1978ev,Senjanovic:1975rk,TypeI-LR}, 
proposed by J. Pati, A. Salam, R. Mohapatra and G. Senjanovi\'c, one can explain some of the open issues of the SM.  In this context, the spontaneous breaking of Parity is naturally explained and one can understand why at the low scale the weak interactions are $V-A$ interactions. These theories predict the existence of 
right-handed neutrinos in nature which play a crucial role to generate neutrino masses. In the context of left-right symmetric theories, neutrinos can be Dirac 
fermions~\cite{Senjanovic:1978ev} or Majorana fermions~\cite{TypeI-LR}. In the Majorana case one can make use of the see-saw mechanism~\cite{TypeI,TypeI-LR} to understand the smallness 
of the neutrino masses. These theories can give rise to many interesting signatures at colliders and low energy experiments, see for example Refs.~\cite{LR-1,LR-2,LR-3,LR-4,LR-5,LR-6,LR-7,LR-8,LR-9,LR-10,LR-11,LR-12,LR-13,LR-14,LR-15,LR-16,LR-17,LR-18,LR-19,LR-20,LR-21,LR-22}
for different phenomenological studies.

Recently, we have proposed a simple left-right symmetric theory~\cite{LRnew} where the Majorana neutrino masses are generated through the Zee-mechanism~\cite{Zee}.
In this context the charged fermion masses are generated at tree level as in the SM, while the neutrino masses are generated at one-loop level. 
This theory has the simplest Higgs sector needed to generate Majorana masses for neutrinos and to realize the spontaneous breaking of the local left-right symmetry obtaining the SM at the weak scale. In Ref.~\cite{LRnew} we have proposed this new theory and investigated the main collider signatures which can help us to test the theory at the Large Hadron Collider.

In this article we investigate in details the properties of the right-handed neutrinos in the theory proposed in Ref.~\cite{LRnew} and the predictions for lepton flavor violating (LFV) processes such 
as the rare decays $e_i \to e_j \gamma$ and $\mu \to e$ conversion. The current experimental bounds from the LFV experiments provide non-trivial bounds on lepton flavour violating interactions present in different theories (see Ref.~\cite{Bernstein:2013hba} for a review of LFV experiments). The next generation of LFV experiments will set very strong bounds on the branching fractions for these rare decays and we investigate the impact of these results in our 
model, in which one has several contributions to LFV processes:  the interactions between the $W_L^{\pm}$ or $W_R^{\pm}$ with the charged leptons and the neutrinos, and the L-violating Higgs interactions. We show that, in this context, one can have very large contributions to 
LFV processes in agreement with all experimental constraints. Together with the collider signatures studied in Ref.~\cite{LRnew} these results can be used to test this theory in current and future experiments. 

\section{Simple Left-Right Symmetric Theory}
%
Recently, we have proposed in Ref.~\cite{LRnew} a simple left-right symmetric theory based on the gauge symmetry 
$$G_{LR}=SU(3)_C \otimes SU(2)_L \otimes SU(2)_R \otimes U(1)_{B-L}$$
where the Majorana neutrino masses are generated at the quantum level. As in any left-right symmetric theory the matter fields live in the following representations
\begin{align*}
&Q_L = \begin{pmatrix} u_L \\ d_L \end{pmatrix} \sim (3, 2, 1, 1/3) \,, \quad Q_R = \begin{pmatrix} u_R \\ d_R \end{pmatrix} \sim (3,1, 2, 1/3) \,,  \\
&\ell_L = \begin{pmatrix} \nu_L \\ e_L \end{pmatrix} \sim (1, 2, 1, -1) \,, \quad  {\rm{and}} \quad \ell_R = \begin{pmatrix} \nu_R \\ e_R \end{pmatrix} \sim (1, 1, 2, -1).
\end{align*}
and the Higgs sector is composed of four Higgses: a bi-doublet needed to generate charged fermion masses, a charged singlet and two doublets required to break the left-right symmetry and generate Majorana neutrino masses in a minimal way,
\begin{align*}
&\Phi =  \begin{pmatrix} \phi_1^0 && \phi_2^+ \\ \phi_1^- && \phi_2^0 \end{pmatrix}  \sim (1, 2, 2, 0) \,, \quad \delta^+ \sim (1,1,1,2),
\end{align*}
\begin{align*}
&H_L = \begin{pmatrix} h_L^+ \\ h_L^0 \end{pmatrix} \sim (1, 2, 1, 1) \, \quad {\rm{and}} \quad H_R = \begin{pmatrix} h_R^+ \\ h_R^0 \end{pmatrix} \sim (1,1, 2, 1). 
\end{align*}
%
\subsection{Charged Fermion Masses}
As in any left-right symmetric theory the charged fermions acquire mass at tree level once the Higgs bi-doublet gets a vacuum expectation value. Using the interactions
\begin{equation}
-\mathcal{L}_{LR}^D=  \overline{Q}_L \left( Y_1 \Phi + Y_2 \tilde{\Phi} \right) Q_R \ + \  \overline{\ell}_L \left( Y_3 \Phi + Y_4 \tilde{\Phi} \right) \ell_R + \rm{h.c.}\ ,
\label{Dirac}
\end{equation}
where $\tilde{\Phi}=\sigma_2 \Phi^* \sigma_2$, one finds the following charged fermion masses after electroweak symmetry breaking
\begin{eqnarray}
M_U &=& Y_1 v_1 + Y_2 v_2^*\,, \\
M_D &=& Y_1 v_2 + Y_2 v_1^*\,, \\
M_E &=& Y_3 v_2 + Y_4 v_1^*.
\label{charged}
\end{eqnarray}
Here, $v_1$ and $v_2$ are the vacuum expectation values for the fields $\phi_1^0$ and $\phi_2^0$, respectively. 
Notice that the $Y_1$ and $Y_2$ can be written as linear combinations of the mass matrices for the up and down quarks, 
while one has more freedom in the expression for charged lepton masses due to the presence of two Yukawa couplings. 
%
\subsection{Neutrino Masses}
%
In the left-right theory with only three Higgses, $\Phi$, $H_L$ and $H_R$, the total lepton number is conserved after symmetry breaking and neutrinos are Dirac massive fermions with mass given by
\begin{eqnarray}
M_\nu^D &=& Y_3 v_1 + Y_4 v_2^*\,.
\label{Diracneutrino}
\end{eqnarray}
Notice that using the freedom in Eqs.(\ref{charged}) and (\ref{Diracneutrino}) one can have a consistent scenario for Dirac neutrinos in this context~\cite{Senjanovic:1978ev}. 
However, the neutrino masses are very small and they could be Majorana fermions. In the theory proposed in Ref.~\cite{LRnew} Majorana neutrino masses are generated at one-loop level using the following interactions
\begin{equation}
- \mathcal{L}_{LR}^M=\lambda_L \ell_L \ell_L \delta^+ \ + \ \lambda_R \ell_R \ell_R \delta^+ + \lambda_1 H_L^T i \sigma_2 \Phi H_R \delta^- + \lambda_2  H_L^T i \sigma_2 \tilde{\Phi} H_R \delta^- + \rm{h.c.}
\label{Majorana}
\end{equation}
See Fig.1 for the one-loop contribution to neutrino masses in the unbroken phase. Notice that both the left-handed and right-handed neutrinos acquire masses at one-loop level 
and their masses are proportional to the vacuum expectation values of $h_L^0$ and $h_R^0$, i.e. $v_L/\sqrt{2}$ and $v_R/\sqrt{2}$. 
\begin{figure}[t]
\includegraphics[width=0.5\linewidth]{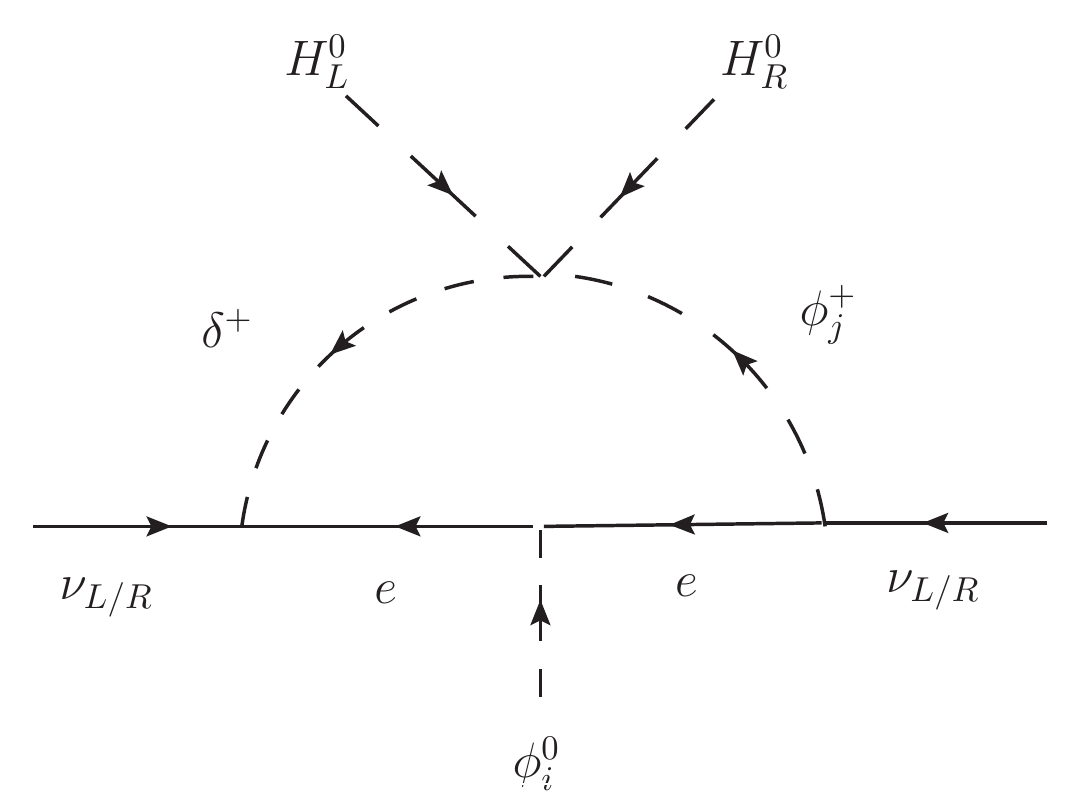}
\caption{Neutrino mass generation at the quantum level.}
\label{Figunbroken}
\end{figure}
The neutrino mass matrix in the basis $\left( \nu_L, \  (\nu^C)_L \right)$ is given by 
\begin{equation}
 \cal{M}_\nu=
\begin{pmatrix} M^L_\nu && M_\nu^D \\ (M_\nu^D)^T && M^R_\nu \end{pmatrix}, 
\end{equation}
where $M^L_\nu$ and $M_\nu^R$ are generated at one-loop level while $M^D_\nu$ is generated at tree level. 
The explicit forms of $M^L_\nu$ and $M_\nu^R$ are given by
\begin{eqnarray}
\displaystyle {(M_\nu^L)}^{\alpha \gamma}&=&\frac{1}{4\pi^2}\lambda_L^{\alpha \beta}m_{e_\beta}\sum_i \text{Log}\left(\frac{M_{h_i}^2}{m_{e_\beta}^2}\right) 
 V_{5i}\left [ (Y_3^\dagger)^{\beta \gamma}V_{2i}^*-(Y_4^\dagger)^{\beta \gamma}V_{1i}^* \right ] \ + \ \alpha \leftrightarrow \gamma 
\,,\\
\displaystyle {(M^R_\nu)}^{\alpha \gamma}&=&\frac{1}{4\pi^2}\lambda_R^{\alpha \beta}m_{e_\beta}\sum_i\text{Log}\left(\frac{M_{h_i}^2}{m_{e_\beta}^2}\right) 
V_{5i}\left[(Y_3)^{\beta \gamma}V_{1i}^*-(Y_4)^{\beta \gamma}V_{2i}^*\right] \ + \ \alpha \leftrightarrow \gamma \,. 
\label{RHM}
\end{eqnarray}
Here $V_{ij}$ defines the mixing between the charged Higgses in the theory and $M_{h_i}$ their 
physical masses (see Ref.~\cite{LRnew} for more details).
In our notation the neutrino mass matrix is diagonalized by the following matrix
\begin{equation}
 \begin{pmatrix} \nu_L \\ (\nu^C)_L\end{pmatrix} \to
\begin{pmatrix} V_\nu && A \\ B && V_N \end{pmatrix}\begin{pmatrix}\nu_L \\ N_L \end{pmatrix}=
\begin{pmatrix} V_\nu \nu_L \,+\,AN_L\\ B\nu_L \,+\, V_N N_L \end{pmatrix},
\label{diagonal_neutrinos}
\end{equation}
which is useful to obtain all physical interactions. Henceforth we are neglecting the mixing 
between the left-handed and right-handed neutrinos because it is very small. 
In general the vevs as well as the Yukawas are free parameters and one cannot predict anything 
about the magnitude of the masses. However, for the theory to be consistent one needs to assume (see 
Ref.~\cite{LRnew}) that $v_2 \ll v_1$ and $Y_3 \ll Y_4$. In this limit, the mass matrix for charged leptons can be approximated by $M_E \approx  Y_4 v_1^*$
and $M^R_\nu$ as 
\begin{equation}
 \displaystyle {(M^R_\nu)}^{\alpha \beta}=\frac{1}{4\pi^2}\lambda_R^{\alpha \beta}\frac{m_{e_\beta}^2}{v_1^*} \sum_i\text{Log}\left(\frac{M_{h_i}^2}{m_{e_\beta}^2}\right)V_{5i}V_{2i}^* \,+\,\alpha \leftrightarrow \beta \,. 
\label{MnuRlim}
\end{equation}
From the above relation one can extract predictions about the hierarchy of the sterile neutrino masses. 
Notice that
Eq.(\ref{MnuRlim}) is traceless due to the product of the antisymmetric Yukawa $\lambda_R^{\alpha \beta}$ with the symmetric mass matrix of the 
charged leptons, which is assumed without loss of generality to be diagonal.
On the other hand, the mass matrix in Eq.~(\ref{MnuRlim}) is sensible to the difference between pairs of charged lepton masses squared. 
Since the difference between the muon mass and the other charged lepton masses is one order of magnitude 
smaller than the rest of differences, the muon neutrino $N_\mu$ is predicted to be at least 
two orders of magnitude lighter than the electron neutrino $N_e$ and the tauon neutrino $N_\tau$.
Bringing together both statements and taking into account the invariance of the trace, the following conclusion about the hierarchy 
of the masses can be drawn: the model predicts that the muon sterile neutrino is much lighter than the others, which therefore have almost degenerated masses.

In order to study qualitatively the order of magnitud of the sterile neutrino masses, one could assume as a good approximation that the product of the 
charged Higgses mixing matrices is of the order of one, due the unitarity nature of the mixing matrix. Notice that unitarity further constraints the sum of the logarithm 
over the different five physical charged Higgses in the theory, making this term only sensible to twice the difference of the order of magnitude between the lightest 
and the heaviest charged Higgses. Let us call this factor $\Delta$, which will represent the contribution of the logarithms in Eq.(\ref{MnuRlim}).
Hence, in the limit $v_2 \ll v_1$ and $Y_3 \ll Y_4$, and assuming that $v_1 \approx v= 246$ GeV,
Eq.~(\ref{MnuRlim}) can be rewritten as
\begin{equation}
 (M_\nu^R)^{\alpha \beta}\approx \frac{\Delta}{4\pi^2 v_1^*}\lambda_R^{\alpha \beta}(m_{e_\alpha}^2 - m_{e_\beta}^2).
 \label{correlation}
\end{equation}
We can estimate a theoretical upper limit for the factor $\Delta$ by assuming the extreme case in which the lightest charged scalar lives at the electroweak scale 
and the heaviest one at the Plank scale. In this scenario $\Delta$ is given by,
%
$ \displaystyle \Delta_{\text{upper}}\sim \text{Log}\left(\frac{10^{19} \text{ GeV}}{10^2 \text{ GeV}}\right)^2 = 2\times 17\times \text{Log}(10)$.
%
The relation in Eq.~(\ref{correlation}) is one of the main predictions of the model, since this correlation between the sterile neutrino masses and the 
Yukawa coupling $\lambda_R$ constrains strongly, on one hand, the hierarchy of the sterile neutrino masses and, on the other hand, allows us to estimate an upper 
bound for the right-handed neutrino masses. 

\begin{figure}[t]
 \includegraphics[width=0.5\linewidth,keepaspectratio=true,clip=true]{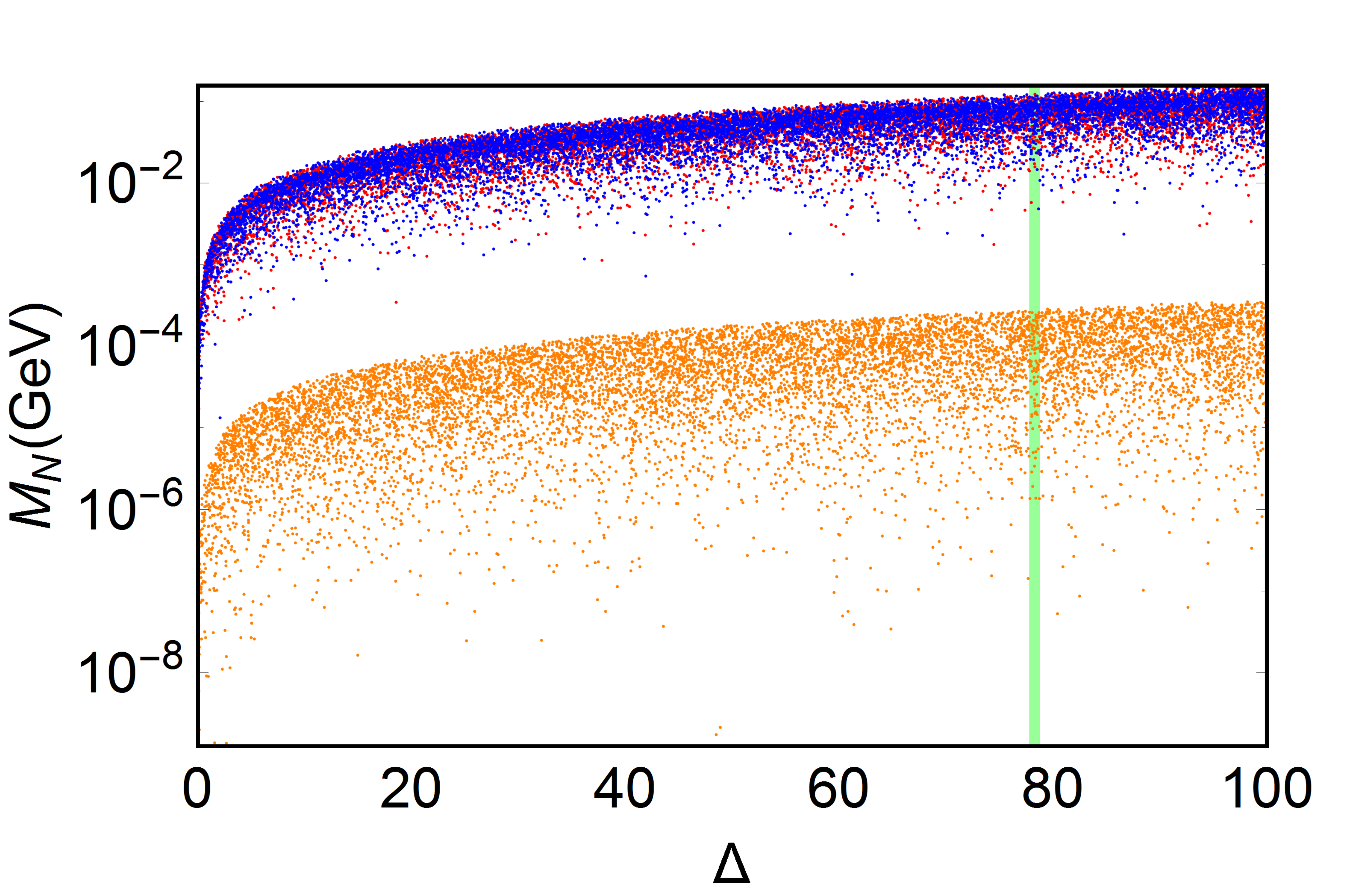}
 \caption{Predictions for the sterile neutrino masses as a function of $\Delta$. The blue, orange and red points correspond to the 
electron $N_e$, muon $N_\mu$, tauon $N_\tau$ sterile neutrinos, respectively. For the scan, the entries of the Yukawa $\lambda_R$ 
have been taken randomly ranging between 0 to $2\sqrt{\pi}$. 
The vertical green line shows the estimated upper bound $\Delta_{\text{upper}}$.}
 \label{neutrino_masses_hierarchy}
\end{figure}

In Fig.~\ref{neutrino_masses_hierarchy} we show the correlation between the sterile neutrino masses $M_{N_i}$, given by the eigenvalues of Eq.~(\ref{correlation}) and the factor $\Delta$. 
The scattered points correspond to different values for the entries of $\lambda_R$, which range randomly from $[0,2\sqrt{\pi}]$, according to perturbativity. As 
we can see in Fig.~\ref{neutrino_masses_hierarchy}, the model predicts very light sterile neutrinos with the theoretically predicted hierarchy. 
We emphasize again the relevance of the antisymmetric nature of the Yukawa $\lambda_R$ matrix since one has only 
three free parameters, $\lambda_R^{e \mu}$, $\lambda_R^{\mu \tau}$ and $\lambda_R^{e \tau}$. Our main result here is that the right-handed neutrinos are generically light. 
This prediction will be very important to study the predictions for lepton flavour violating processes. 
%
\subsection{Charged Gauge Boson Masses}
%
In the basis $(W_L^+,  \ W_R^+)$ the charged gauge boson mass matrix reads as
\begin{equation}
 {\cal{M}}^2_{\pm}=\begin{pmatrix} \frac{g_L^2}{2}(\frac{1}{2}v_L^2 + v^2) && - g_L g_R v_1 v_2 \\ - g_L g_R v_1 v_2 && \frac{g_R^2}{2}(\frac{1}{2}v_R^2 + v^2) \end{pmatrix},
\end{equation}
where $v^2 = v_1^2 + v_2^2$. The mass of the $W_R$-like charged gauge boson is $M_{W_R} \approx g_R v_R /2$. 
Using the charged current interactions 
\begin{equation}
 \displaystyle -{\cal L}^{CC}_{LR}\supset \frac{g_L}{\sqrt{2}} \bar{\nu}_L \slashed{W}^+_L e_L + \frac{g_R}{\sqrt{2}}
  \bar{\nu}_R \slashed{W}^+_R e_R + \text{h.c.} \ ,
\end{equation}
and the definition in Eq.(\ref{diagonal_neutrinos}) one can study lepton flavour violation in the leptonic sector mediated by the gauge bosons. 
Recently, the LHC experiments have set bounds on the mass of these gauge bosons. See Ref.~\cite{Khachatryan:2016jww} for the lower 
experimental bound, $M_{W_R} > 4.1$ TeV, on the mass of the $W_R$-like gauge boson.
%
\section{Lepton Flavour Violating Processes}
%
In the theory proposed in Ref.~\cite{LRnew} there are several sources of lepton flavour violation:
\begin{itemize}

\item The physical interactions between the $W_L^{\pm}$, the charged leptons and the neutrinos. 

\item The physical interactions between the $W_R^{\pm}$, the charged leptons and the neutrinos.

\item In Eq.(\ref{Dirac}) we cannot simultaneously diagonalize the Yukawa couplings $Y_3$ and $Y_4$, and the Higgs interactions violate the family lepton numbers.  

\item The Yukawa interactions in Eq.(\ref{Majorana}) violate the global $U(1)_{L_i}$ as well.

\end{itemize} 
In this section we investigate the predictions for lepton flavour violating processes such as $e_i \to e_j \gamma$ taking into account the different sources for $L_i$ violation. 
Notice that, among the different sources violating lepton flavour, there are the charged Higgses in the bi-doublet. In this work we will not focus on them since 
they also contribute to $\Delta F=2$ hadronic changing neutral current effects which are very constrained and, therefore, these Higgses have to be 
very heavy~\cite{LR-4}. However, in the context of the recently proposed left-right symmetric model~\cite{LRnew}, the singly charged Higgs could be relatively light 
and can induce large contributions to lepton flavour violating processes.
%
\subsection{LFV \texorpdfstring{$e_i \to e_j \gamma$}{ei -> ej gamma} Processes}
\begin{figure}[t]
\includegraphics[width=0.8\linewidth,keepaspectratio=true,clip=true]{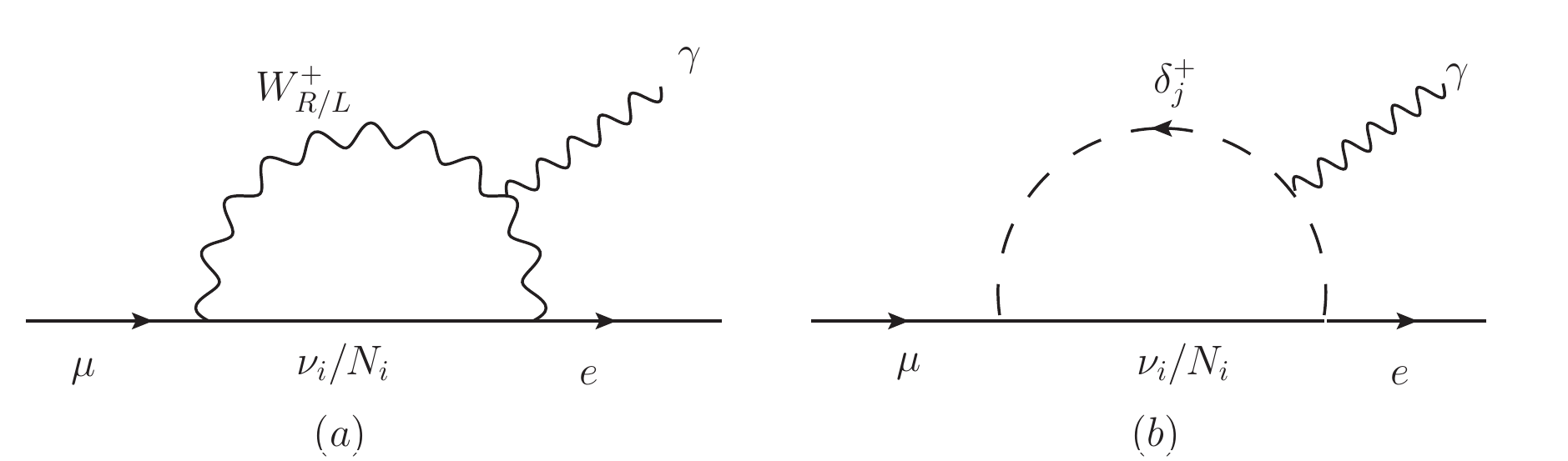}
\caption{Contributions to $\mu \to e \gamma$. Topology (a) mediated by a charged gauge boson, $W_L^\pm$ or $W_R^\pm$, and topology (b) mediated by a singly charged Higgs.}
\label{mutoemediators}
\end{figure}

In this section we investigate the predictions for the lepton flavour violating processes in order to understand the testability of the theory 
proposed in Ref.~\cite{LRnew}. The current experimental bounds on the branching ratios for the $e_i \to e_j \gamma$ processes are
\begin{equation*}
\rm{Br}(\mu \to e \gamma)< 4.2 \times 10^{-13} \mbox{\cite{Adam:2013mnn}}, \ \rm{Br} (\tau \to e \gamma) < 3.3 \times 10^{-8} \mbox{\cite{Aubert:2009ag}}, \
\rm{Br} (\tau \to \mu \gamma) < 4.4 \times 10^{-8} \mbox{\cite{Aubert:2009ag}}.
\end{equation*}
As it is well known, the amplitude for the process $\mu \to e \gamma$ can be written as
\begin{equation}
  \mathcal{A}(\mu\to e \gamma)=i\overline{u_e}(p-q)\epsilon^*_{\nu}\sigma^{\nu \mu}q_{\mu}[A_RP_R+A_LP_L]u_{\mu}(p),
\end{equation}
where $p^{\mu}$ and $q^\mu$ are the muon and photon quadrimomenta, respectively, and the decay width reads as
\begin{equation}
 \displaystyle \Gamma(\mu \to e\gamma)=\frac{m_\mu^3}{16\pi}(|A_L|^2+|A_R|^2).
 \label{decaywidth}
\end{equation}
The branching ratio can be computed using the relation
\begin{equation}
 \text{Br}(\mu \to e \gamma)\equiv \frac{\Gamma(\mu \to e \gamma)}{\Gamma(\mu \to e \nu_\mu\bar{\nu_e})+\Gamma(\mu \to e \gamma)},
\end{equation}
where
\begin{equation}
 \Gamma(\mu \to e \nu_\mu \bar{\nu_e})=\frac{m_\mu^5G_F^2}{192\pi^3}.
\end{equation}
In our model there are several contributions to the coefficients $A_L$ and $A_R$ relevant for the decay width. 
In Fig.~\ref{mutoemediators} we show the Feynman graphs for the different contributions mediated by the charged gauge bosons and the charged Higgses.
Here we will investigate the predictions for each contribution in order to understand the testability of the theory in current and future experiments.
%
\subsubsection{LFV induced by $W^{\pm}_i$ gauge bosons}
\begin{figure}[t]
 \includegraphics[width=0.45\linewidth,keepaspectratio=true,clip=true]{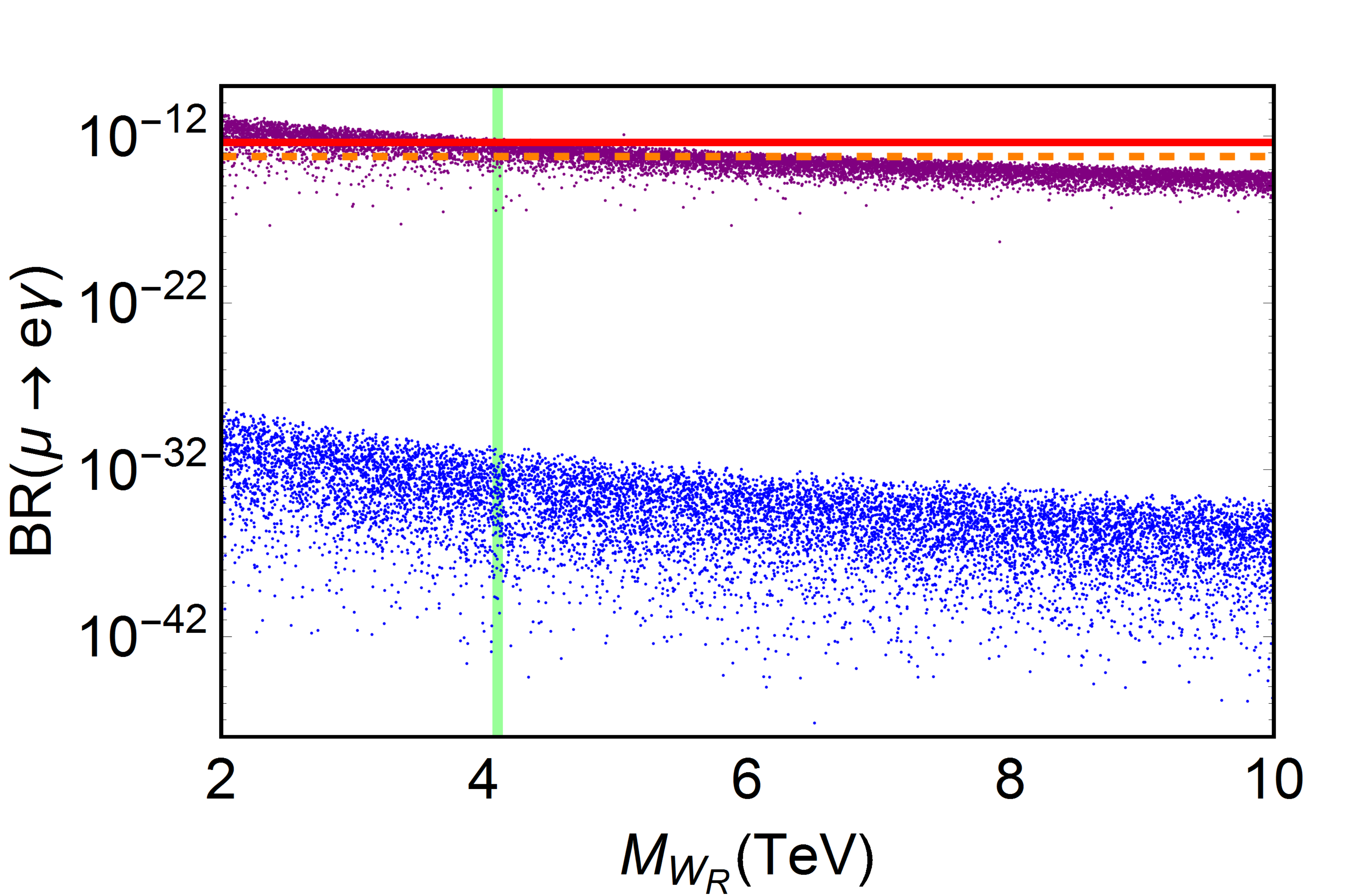}
 \includegraphics[width=0.45\linewidth,keepaspectratio=true,clip=true]{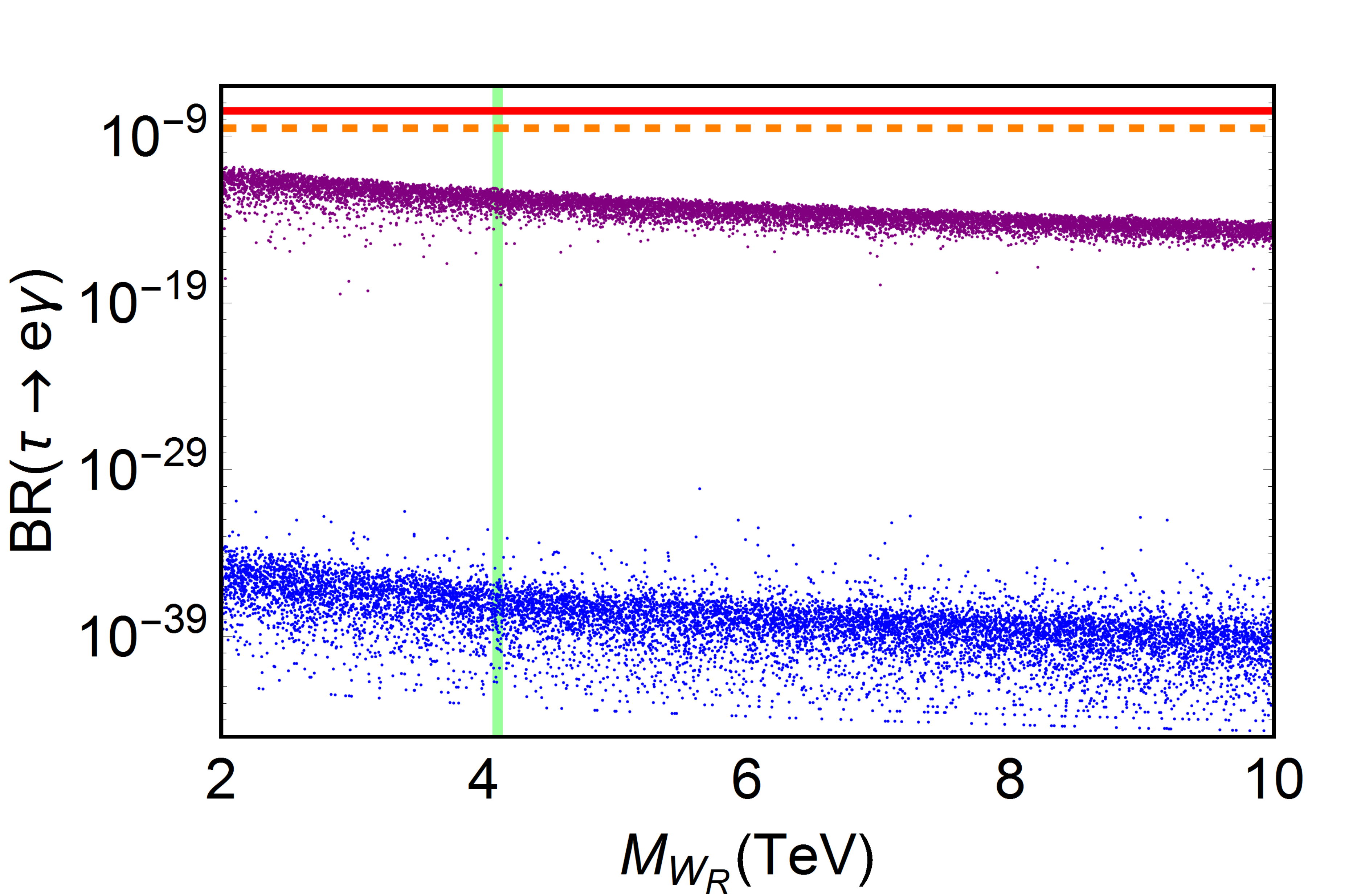}
 \includegraphics[width=0.45\linewidth,keepaspectratio=true,clip=true]{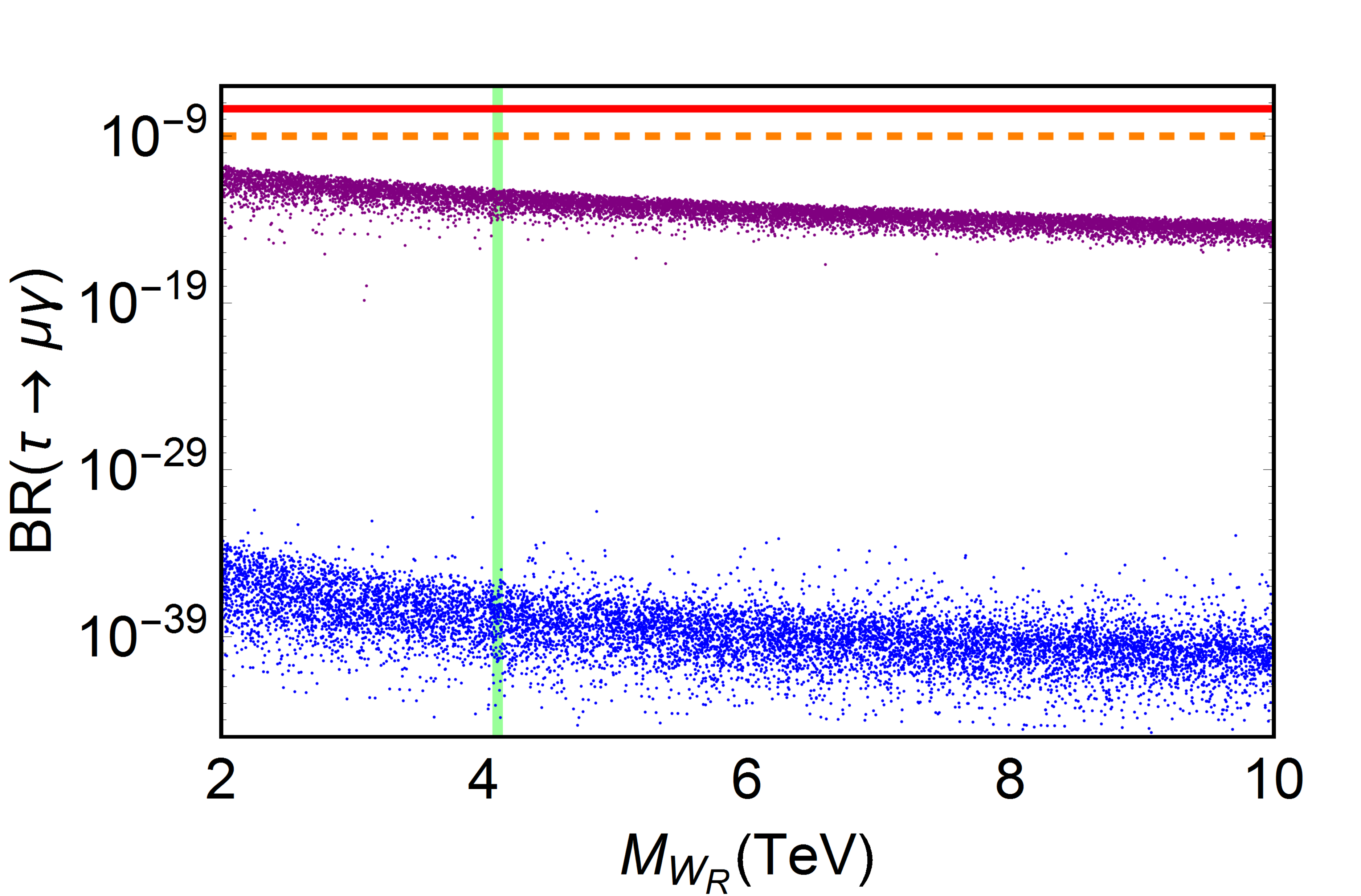}
 \includegraphics[width=0.45\linewidth,keepaspectratio=true,clip=true]{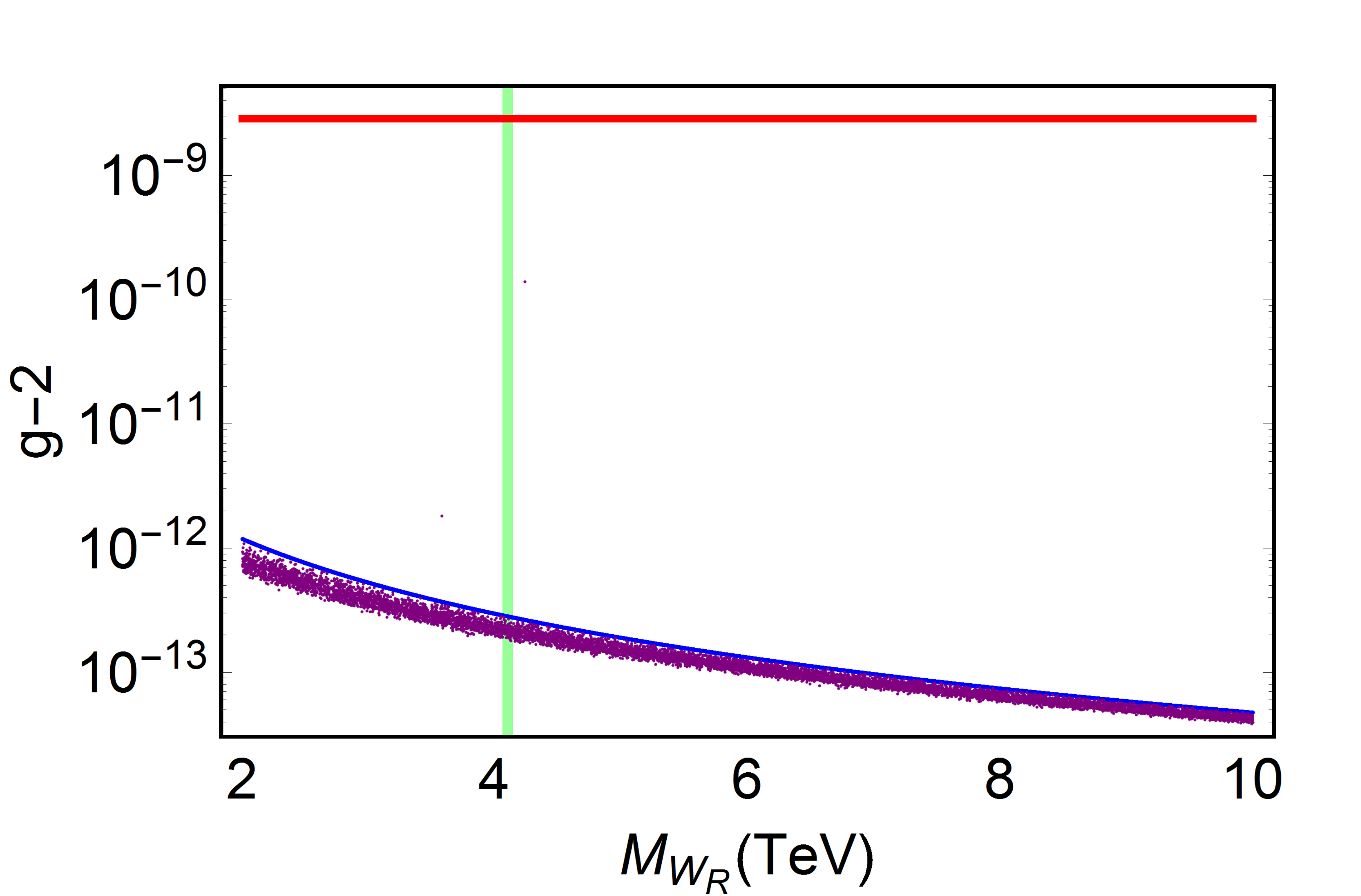}
 \caption{Predictions on the branching ratio of $e_i\to e_j \gamma$ and $g-2$ anomalous momentum of the muon mediated by a right handed charged gauge boson 
 as a function of its mass $M_{W_R}$. The blue points represent the predictions on the branching ratio in the context of our model, whereas the purple points 
 correspond to the most optimistic case to observe LNV in which the right handed neutrinos and the $W_R$ boson masses are of the same order of magnitude. For 
 all points, $\lambda_R \in [0,2\sqrt{\pi}]$ (perturbative scenario) and the factor $\Delta \in [0,\Delta_{\text{upper}}]$. 
 The red lines represent the experimental lower bounds on the different processes, $\rm{Br}(\mu \to e \gamma)< 4.2 \times 10^{-13}$ \cite{Adam:2013mnn}, 
$ \rm{Br} (\tau \to e \gamma) < 3.3 \times 10^{-8}$ \cite{Aubert:2009ag}, $\rm{Br} (\tau \to \mu \gamma) < 4.4 \times 10^{-8}$ \cite{Aubert:2009ag} and $\Delta a_\mu < 287 \times 10^{-11}$ \cite{Blum:2013xva}. The 
 orange dashed lines represent the projected limits $6\times 10^{-14}$ for $\mu \to e \gamma$ \cite{Baldini:2013ke}, $3\times 10^{-9}$ for $\mu \to e\gamma$ \cite{Aushev:2010bq}, 
 and $\sim 10^{-9}$ for $\tau \to e\gamma$ \cite{Aushev:2010bq}. Finally, the green vertical line shows the lower bound on the $W_R$ gauge boson \cite{Khachatryan:2016jww}, which only applies to the blue points.
  Here, left-right symmetry has been assummed in the gauge sector, i.e. $g_L=g_R$.}
 \label{scan_WR}
\end{figure}

%
The $W_L^{\pm}$ and $W_R^{\pm}$ contributions to the process $\mu \to e \gamma$, neglecting the electron mass, read as 
 \begin{eqnarray}
  A_L^{W_R}&\approx& g_R^2\frac{e \ m_\mu}{64 \pi^2 M_{W_R}^2}\sum_i(V_N)_{e i}(V_N^*)_{\mu i}F\left(\frac{m_{N_i}^2}{m_{W_R}^2}\right),
\end{eqnarray}  
\begin{eqnarray}  
  A_R^{W_L}&\approx& g_L^2\frac{e \ m_\mu}{64 \pi^2 M_{W_L}^2}\sum_i(V_\nu)_{e i}(V_\nu^*)_{\mu i}F\left(\frac{m_{\nu_i}^2}{m_{W_L}^2}\right),
 \end{eqnarray}
where we have neglected the mixing between the neutrinos. The loop scalar function $F(x)$ is defined as
\begin{equation}
 \displaystyle F(x)=\frac{1}{6(1-x)^4}\left(10-43\,x+78\,x^2-49\,x^3+18\,x^3\, \text{Log}(x)+4\,x^4\right),
\end{equation}
which has the following limits,
\begin{equation}
F(x)_{x\to \infty} \sim \frac{2}{3}+3\ \frac{\text{Log} (x)}{x},\quad
F(x)_{x\to 0} \sim \frac{5}{3}-\frac{1}{2}x,\quad
F(x)_{x \to 1} \sim \frac{17}{12}+\frac{3}{20}(1-x).
\end{equation}
Notice that when $W_L^\pm$ is the mediator of the process, the Standard Model neutrinos are the ones contributing into the amplitude. In the second contribution, the $W_R^\pm$ and 
the sterile neutrinos are inside the loop. In the case of $W_L^\pm$, $x\equiv m_{\nu_i}^2/M_{W_L}^2\to 0$ due to the smallness of the Standard Model neutrino masses and the function $F(x\to 0)$ 
characterizing the loop behaves almost as a constant. Hence, $A_R^{W_L}\sim 0$ due to the unitarity of $V_\nu$, i.e. the so-called GIM suppression.
For the $W_R$ gauge boson, however, the GIM suppression can be avoided if the sterile neutrinos are heavy enough to spoil the suppression coming from the unitarity relations. 
Each of the limits of $F(x)$ leads to a different amplitude, shown in Table~\ref{scenarios}.
\begin{table}[h] 
  \caption{Different limits for $A_L^{W_R}$ amplitud.} 
  \vspace{0.5cm}
  \centering
 \begin{tabular}{ c | c }
   \hline
  \multicolumn{2}{l}{\,\,\,\,\,\,\,\,\,\,Limit \,\,\,\,\,\,\,\,\,\,\,\,\,
  \,\,\,\,\,\,\,\,\,\,\,\,\,\,\,\,\,\,\,\,\,\,\,\,\,\,\,\,\,\,\,\,\,\,\,\,\,\,\,\,\,\,\,\,\,\,\,\,\,\,$A_L^{W_R}$} \\
  \hline
   $m_{N_i}\ll M_{W_R}$ & $ -g_R^2\frac{ e}{128 \pi^2}\frac{m_\mu}{M_{W_R}^4}\sum_i (V_N)_{e i}(V_N^*)_{\mu i} m_{N_i}^2$ \\  
 
  $m_{N_i}\gg M_{W_R}$ & $  g_R^2m_\mu\frac{3 e}{64\pi^2}\sum_i\text{Log}\left(\frac{m_{N_i}^2}{M_{W_R}^2}\right)(V_N)_{i e}(V_N^*)_{i \mu}\frac{1}{m_{N_i}^2}$ \\ 
  
  $m_{N_i} \sim M_{W_R}$ & $ g_R^2\frac{3}{1280 \pi^2}\frac{m_\mu}{M_{W_R}^2}\sum_i (V_N)_{e i}(V_N^*)_{\mu i}\frac{M_{W_R}^2-m_{N_i}^2}{M_{W_R}^2}$ \\
  \hline
  \end{tabular}
 \label{scenarios}
 \end {table}
The largest possible contribution corresponds to the case where both, the masses of the sterile neutrinos and the mass of the charged gauge boson, are of the same order of 
magnitude, i.e. $m_{N_i} \sim M_{W_R}$. We illustrate this scenario in Fig.~\ref{scan_WR} by showing the prediction on the branching ratio 
$e_i \to e_j \gamma$ (purple points) as a function of $M_{W_R}$ for different values of the sterile neutrino masses of the same order of magnitude as $M_{W_R}$.
%
%
In the context of our LR-model in Ref.~\cite{LRnew}, neutrinos are predicted to be very light;  $m_{N_i}/ M_{W_R} \leq 10^{-4}$. 
Therefore, as we also show in Fig.~\ref{scan_WR}, blue points, the GIM suppression occurs and the predictions for the 
branching ratio are far away from the current and even future experimental reach. In order to complete our discussion we show the predictions 
for $g-2$ which are always very small.
%
\subsubsection{LFV mediated by $\delta^{\pm}$ Charged Higgses}
%
The charged Higgs, $\delta^{\pm}$, can be light and induce large contributions to the lepton flavour violating processes.
The amplitude for the $\mu \to e \gamma$ process can be written as
\begin{eqnarray}
 A_L^{\delta^+} &=& \frac{e}{4\pi^2}\frac{m_\mu}{m_{\delta^+}^2}\sum_i\sum_{c,d}(\lambda_R^*)^{ce}\lambda_R^{d\mu}V_N^{ci}(V_N^*)^{di}\,G\left(
 \frac{m_{N_i}^2}{m_{\delta^+}^2}\right), \label{ARS}\\
 A_R^{\delta^+} &=& \frac{e}{4\pi^2}\frac{m_\mu}{m_{\delta^+}^2}\sum_i\sum_{c,d}(\lambda_L^*)^{ce}\lambda_L^{d\mu}(V_\nu^*)^{ci}V_\nu^{di}\,G\left(
 \frac{m_{\nu_i}^2}{m_{\delta^+}^2}\right),  \label{ALS}
\end{eqnarray}
where, as in previous discussion, fermion masses in the loop have been neglected, and the scalar function $G(x)$ is defined as
\begin{equation}
 G(x)=\frac{1-6x+3x^2+2x^3-6x^2\text{Log}(x)}{12(1-x)^4}.
 \label{G}
\end{equation}
Notice that, as one can see in Fig.~\ref{mutoemediators} (b) as well as in the above equation, the $N_i$ and $\delta^+$ decouple in the limit where their masses are heavy so that the amplitude contributing to LFV is suppressed. Therefore, only when both masses are light, LFV processes can have a large effect. 
As we have already commented, the crucial point here is that $\lambda_R$ has flavour indices and therefore breaks the unitarity relations. 
However, LNV processes not protected by the GIM suppression or any given internal symmetry are dangerous in the sense that they could easily be in conflict with the 
strong current experimental bounds. It is remarkable that, despite of losing the GIM protection, the consistency of the left-right symmetric model with a charged scalar introduced in 
Ref.~\cite{LRnew} is ensured due to the light sterile neutrinos that this theory predicts. In this article we neglect the mixing between $\delta^+$ and the other charged Higgses because it is always small. The term in the scalar potential which define this mixing is $H_L^T i \sigma_2 \Phi H_R \delta^-$, and since the charged Higgses in the bidoublet 
are very heavy, the mixing angle is always very small.  

\begin{figure}[t]
 \includegraphics[width=0.45\linewidth,keepaspectratio=true,clip=true]{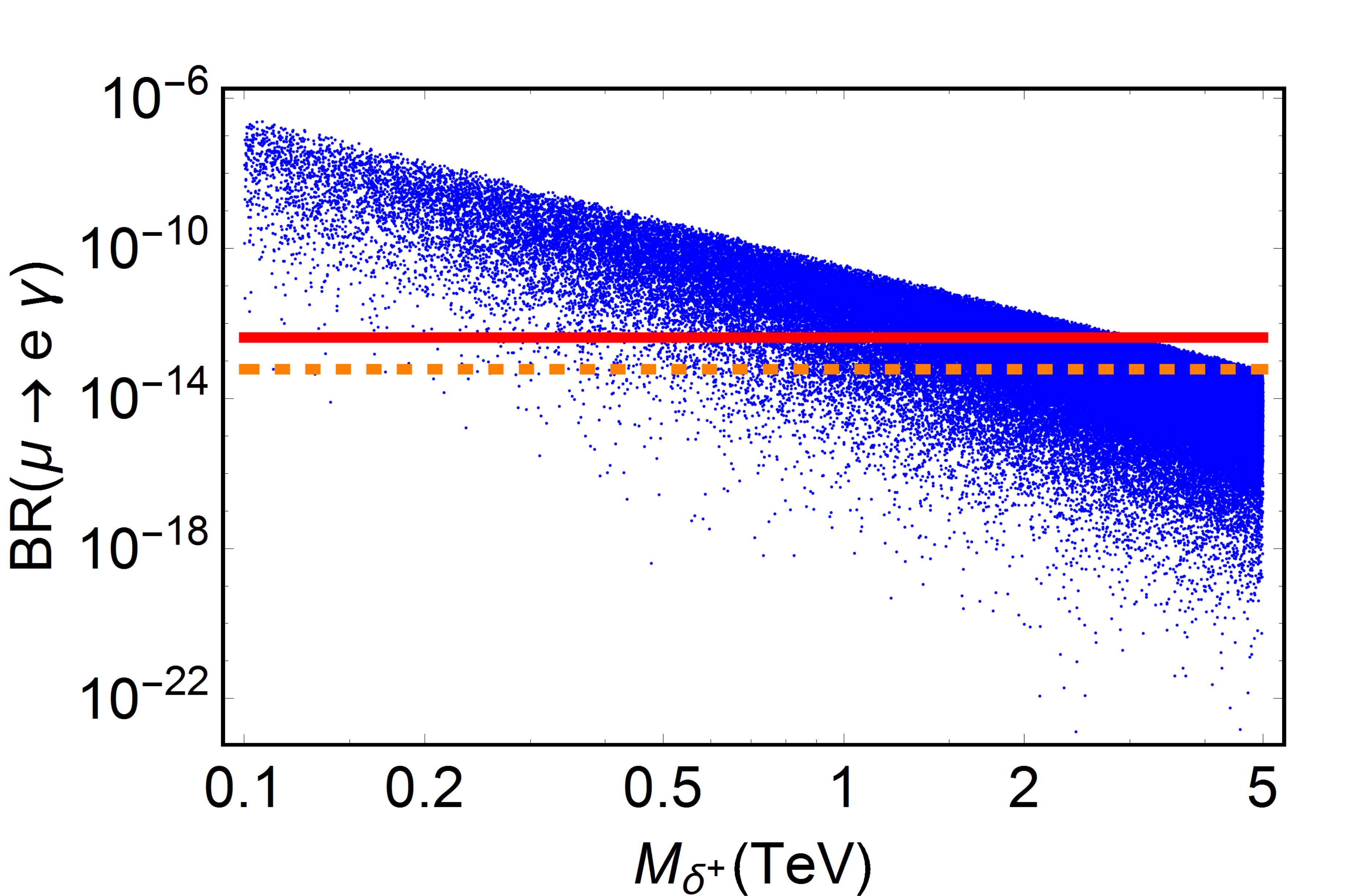}
 \includegraphics[width=0.45\linewidth,keepaspectratio=true,clip=true]{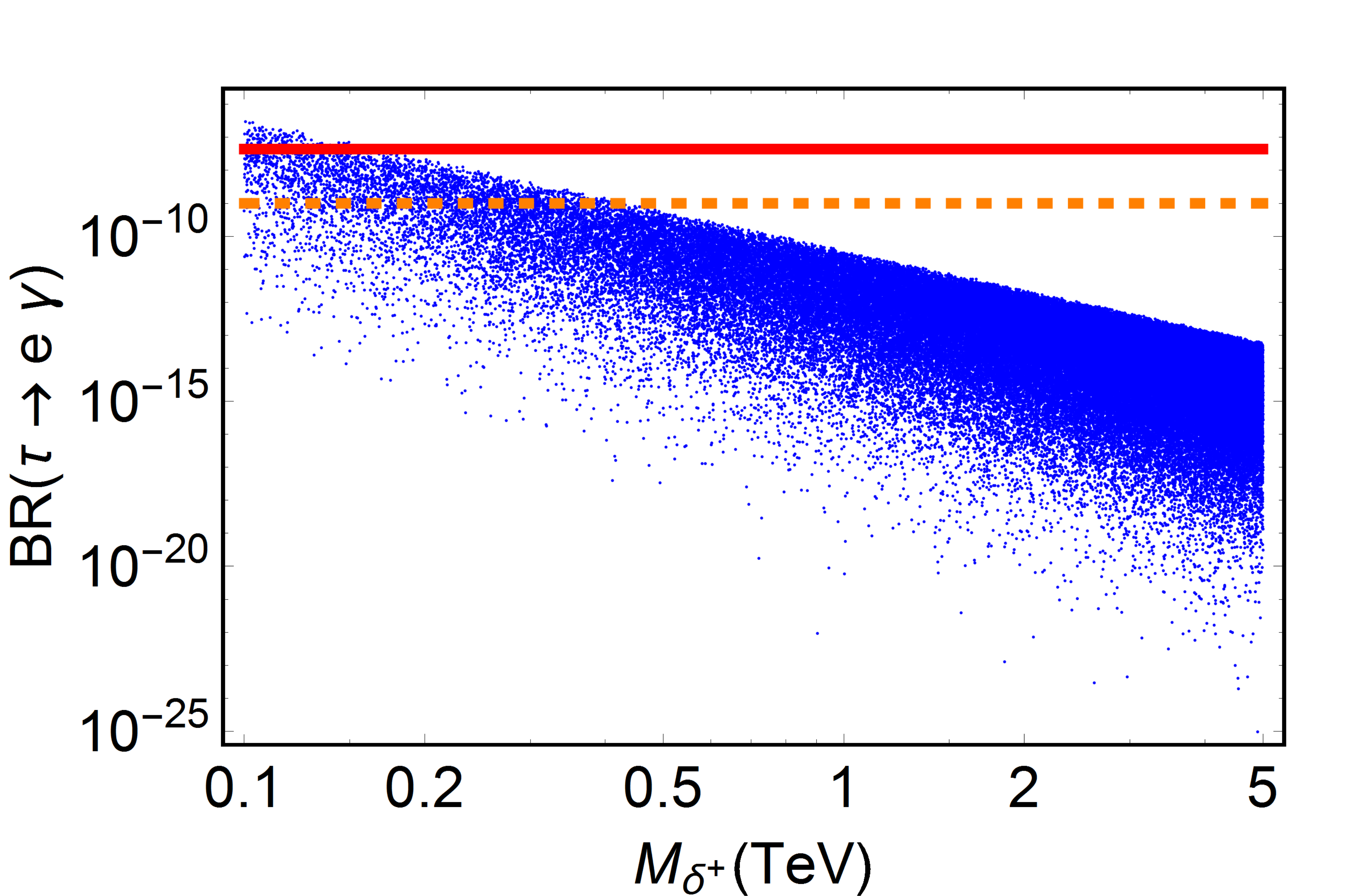}
 \includegraphics[width=0.45\linewidth,keepaspectratio=true,clip=true]{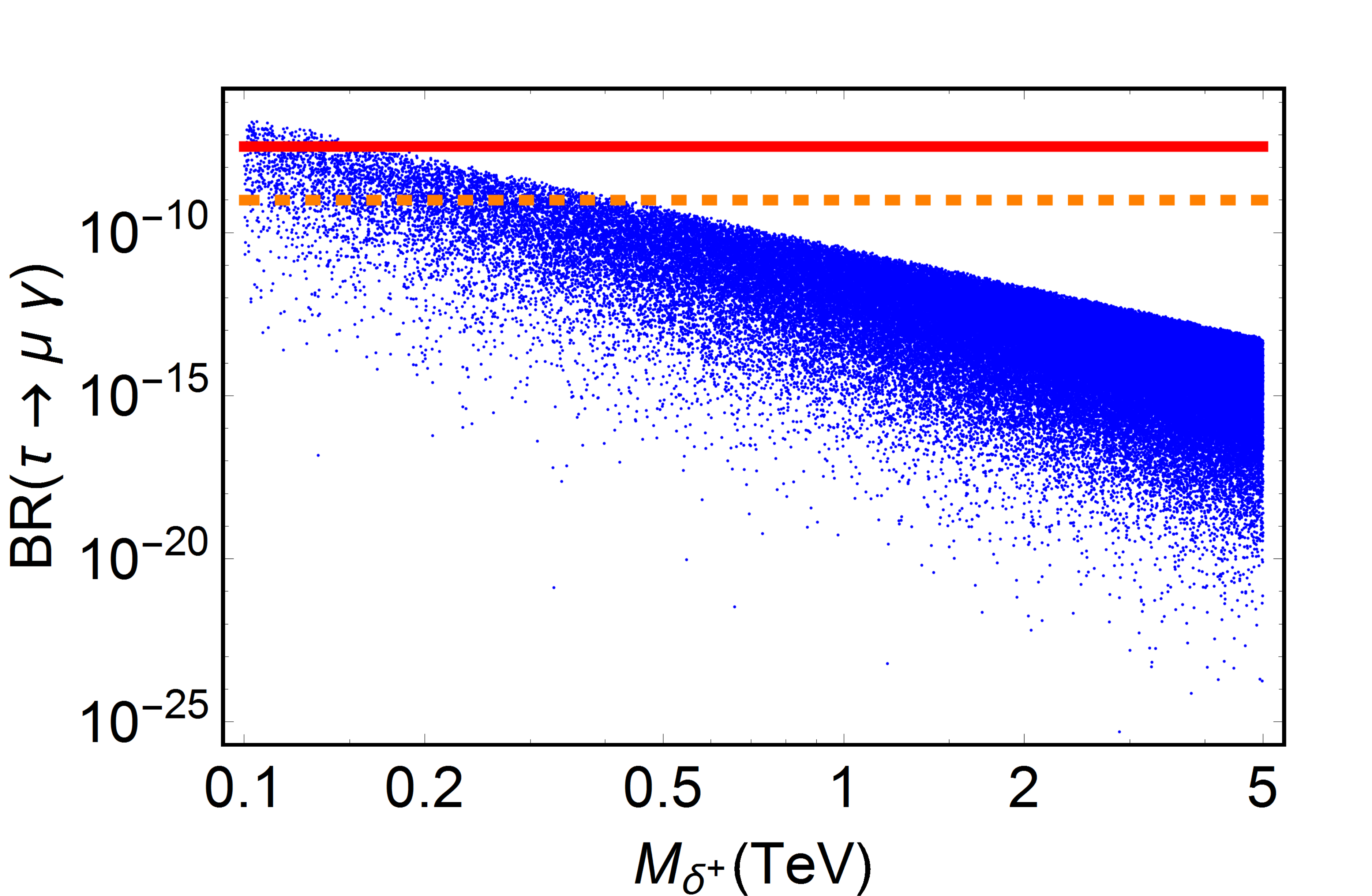}
 \includegraphics[width=0.45\linewidth,keepaspectratio=true,clip=true]{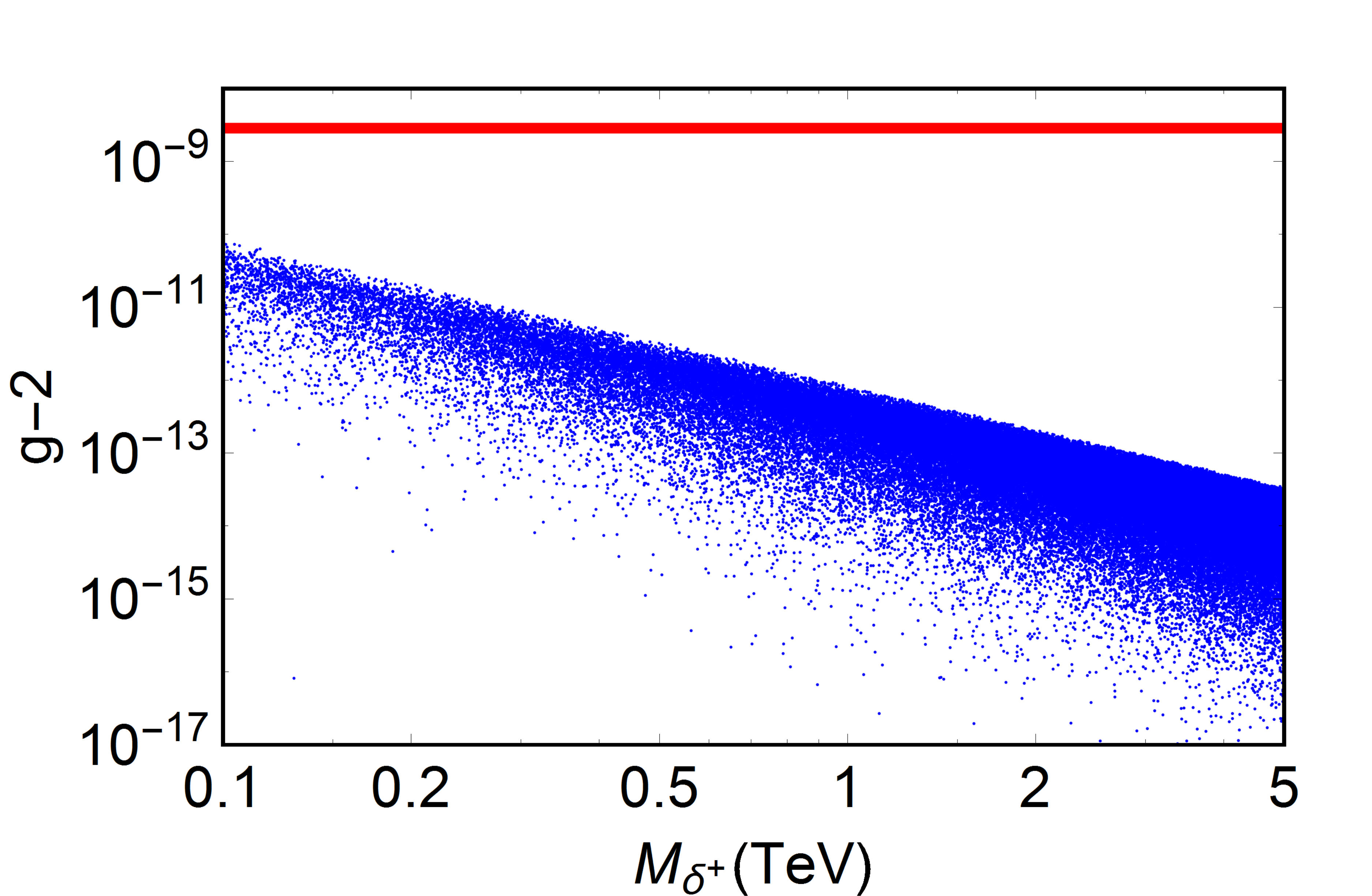}
 \caption{Predictions on the branching ratio of the process $e_i \to e_j \gamma$ mediated by a singly charged scalar 
 as a function of the $M_{\delta^+}$. Here, the Yukawa $\lambda_R \in [0,0.1]$ and the factor $\Delta \in [0,\Delta_{\text{upper}}]$. 
 The red lines represent the experimental lower bounds on the different processes, $\rm{Br}(\mu \to e \gamma)< 4.2 \times 10^{-13}$ \cite{Adam:2013mnn}, 
 $\rm{Br} (\tau \to e \gamma) < 3.3 \times 10^{-8}$ \cite{Aubert:2009ag}, $\rm{Br} (\tau \to \mu \gamma) < 4.4 \times 10^{-8}$ \cite{Aubert:2009ag} and $\Delta a_\mu < 287 \times 10^{-11}$ \cite{Blum:2013xva}. The 
 orange dashed lines represent the projected limits $6\times 10^{-14}$ for $\mu \to e \gamma$ \cite{Baldini:2013ke}, $3\times 10^{-9}$ for $\mu \to e\gamma$ \cite{Aushev:2010bq}, 
 and $\sim 10^{-9}$ for $\tau \to e\gamma$ \cite{Aushev:2010bq}.}
 \label{scanS}
\end{figure}
In Fig.~\ref{scanS} we show the predictions on the branching ratio for $e_i\to e_j \gamma$ mediated by $\delta^+$ as a function of the mass of the charged scalar. The points plotted range from $\lambda_R\in[0,0.1]$, i.e. the Yukawa coupling has been assumed to be perturbative, and the $\Delta$ factor from $\Delta \in [0,\Delta_{\text{upper}}]$. Notice that, due to the relation shown in Eq.~(\ref{correlation}), the sterile neutrino masses as well as the rotation matrix $V_N$ can be extracted from the coupling $\lambda_R$ and the charged fermion masses. 
We neglected the contributions mediated by the charged gauge bosons since they are very small. In Fig.~\ref{scanS} one can see that the limits on the branching ratio 
for the process $\mu \to e \gamma$ impose non-trivial bounds on the mass of the charged Higgs generating neutrino masses at the one-loop level. However, 
generically the charged Higgs $\delta^{\pm}$ can be light enough to avoid the LFV bounds and be produced at the Large Hadron Collider with a large cross section. In Fig.~\ref{scanS} 
we also show the numerical results for $g-2$, which are very small, in order to complete our discussion. 
\FloatBarrier
%
\subsection{\texorpdfstring{$\mu \to e $}{mu-e} conversion}
%
The process of $\mu \to e$ conversion was first studied by Weinberg and Feinberg \cite{Weinberg:1959zz}. 
See also Refs.~\cite{Marciano:1977cj,Czarnecki:1998iz,Cirigliano:2009bz,Kitano:2002mt,Cirigliano:2004mv}.
The experimental lower bounds for this process are stronger than for the $\mu \to e \gamma$ process
%
$$R_{\mu \to e}^{\text{Ti}} < 4.3 \times 10^{-12} \mbox{\cite{Dohmen:1993mp}}, R_{\mu \to e}^{\text{Au}} < 7 \times 10^{-13} \mbox{\cite{Bertl:2006up}}, R_{\mu \to e}^{\text{Pb}}< 4.6\times 10^{-11} \mbox{\cite{Honecker:1996zf}},$$
%
and, moreover, these constraints are expected to be improved by several orders of magnitude in a near future according to some future experiments such as 
DeeMe at J-PARC \cite{Aoki:2010zz}, with an expected sensitivity of $10^{-14}$, Mu2e at Fermilab \cite{Morescalchi:2016uks}, with $6\times 10^{-17}$, and COMET at J-PARC \cite{Cui:2009zz}, with $10^{-16}$. These competitive lower bounds make the process $\mu-e$ conversion very attractive to constrain models predicting lepton flavour violating interactions.

\begin{figure}[t]
 \includegraphics[width=0.8\linewidth,keepaspectratio=true,clip=true]{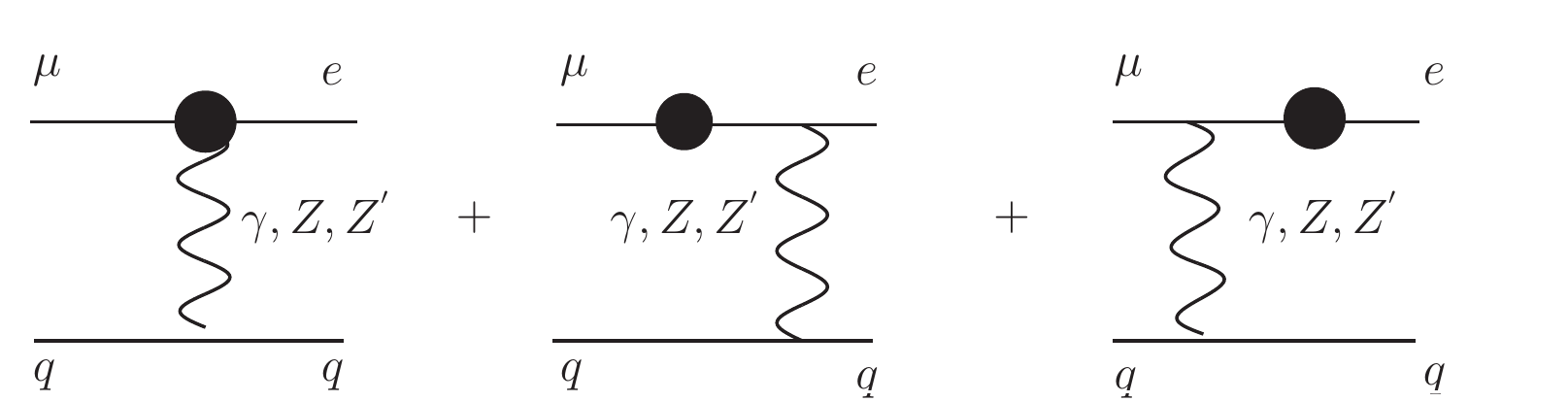}
 \caption{Diagramatic contributions to $\mu$ to $e$ conversion. The black spot symbolizes the effective vertex.}
 \label{mutoe}
\end{figure}

From an effective theory point of view, the effective interactions contributing to this process in the context of our left-right symmetric model with 
a singly charged Higgs can be written as
\begin{equation}
- {\cal L}_{\text{eff}}^{(q)}= C_{DR}\,m_\mu \, \bar{e}\sigma^{\rho \nu}P_L \mu \, F_{\rho \nu}+\sum_qC_{VR}^{(q)}\bar{e}\gamma^\rho P_R \mu (\bar{q} \gamma_\rho q)
 +\text{h.c.}
 \label{effective_lagrangian}
\end{equation}
Among the gauge boson contributions to the above lagrangian, shown in Fig.~\ref{mutoe}, we only consider the photonic contribution since the contribution of the 
massive gauge bosons is suppressed by their masses squared. As a good approximation, we assume that the transferred momentum $t$ of the process shown in Fig.~\ref{mutoe} 
is of the order of $t\sim m_\mu^2$, which would correspond to an elastic collision. The Wilson coefficients in Eq.(\ref{effective_lagrangian}) in the context of our 
left-right symmetric model are written explicitly in the appendix. Notice that here, unlike in the $\mu \to e \gamma$ case, the photon is off-shell. However, since $t\sim m_\mu^2$, the transferred 
momentum can be neglected inside the loop compared to the charged scalar/gauge bosons masses, 
giving the same result as the one showed in the section of $\mu \to e \gamma$. Here the contributions mediated by the charged 
gauge bosons have been neglected in front of the contribution of the singly charged Higgs, as we have justified before.
The above lagrangian is defined at the quark level. However, we are interested in low energy processes involving nuclei. In order to perform the matching between both 
energy regimes we follow Refs.~\cite{Cirigliano:2009bz} and \cite{Kitano:2002mt}, where the matching between quark and parton level is made by introducing the following 
form factors
\begin{equation}
 \bar{q}\gamma_{\rho}q\rightarrow f_{VN}^{(q)}\overline{\Psi}_N\gamma_{\rho}\Psi_N,
\end{equation}
defined as
\begin{eqnarray}
 f_{Vp}^{(u)}=2, \,\,\,\,\, f_{Vp}^{(d)}=1, \,\,\,\,\, f_{Vp}^{(s)}=0,\\
 f_{Vn}^{(u)}=1, \,\,\,\,\, f_{Vn}^{(d)}=2, \,\,\,\,\, f_{Vn}^{(s)}=0.
\end{eqnarray}
and thus the new effective couplings $\tilde{C}_{VR}^{(N)}$ read as
\begin{eqnarray}
\tilde{C}_{VR}^{(p)}&=&\sum_{q=u,d,s}C_{VR}^{(q)}f_{Vp}^{(q)},\\
\tilde{C}_{VR}^{(n)}&=&\sum_{q=u,d,s}C_{VR}^{(q)}f_{Vn}^{(q)}.
\end{eqnarray}
The muon conversion rate reads as~\cite{Kitano:2002mt}
\begin{equation}
 \Gamma_{\text{conv}}=\frac{m_\mu^5}{4}\left |C_{DR}\, D+\tilde{C}_{VR}^{(p)}4V^{(p)}+\tilde{C}_{VR}^{(n)}4V^{(n)}\right |^2,
\label{conversion_rate}
\end{equation}
where the dimensionless integrals $D$ and $V^{(N)}$ represent the overlap of electron and muon wavefunctions and depend on the nucleus involved. Their explicit expressions 
are given by~\cite{Kitano:2002mt}
\begin{equation}
 D=\frac{4}{\sqrt{2}}\, m_\mu \int_0^{\infty}{dr\,r^2[-E(r)](g_e^-f_\mu^-+f_e^-g_\mu^-)},
\end{equation}
\begin{eqnarray} 
 V^{(p)}&=&\frac{1}{2\sqrt{2}}\int_0^{\infty}{dr\,r^2Z\rho^{(p)}(g_e^-g_\mu^-+f_e^-f_\mu^-)},\\
 V^{(n)}&=&\frac{1}{2\sqrt{2}}\int_0^{\infty}{dr\,r^2(A-Z)\rho^{(n)}(g_e^-g_\mu^-+f_e^-f_\mu^-)},
\end{eqnarray}
where $\rho^{(N)}$ is the density of the nucleon, 
$E(r)$ refers to the electric field, which is obtained by integrating the Maxwell equation
\begin{equation}
 E(r)=\frac{Ze}{r^2}\int_0^r{r^{'2}\rho^{(p)}(r^{'})dr^{'}},
\end{equation}
and the functions $g_\mu^-$, $f_\mu^-$, $g_e^-$ and $f_e^-$ correspond to the 1s muon wave functions and electron wave functions, respectively, according to the nomenclature used in Ref.~\cite{Czarnecki:1998iz}. 
The values of the dimensionless integrals $D$ and $V^{(N)}$ along with the capture rates 
of the named isotopes are listed in Table~\ref{Tab_mutoe}.
\begin{figure}[t]
 \includegraphics[width=0.45\linewidth,keepaspectratio=true,clip=true]{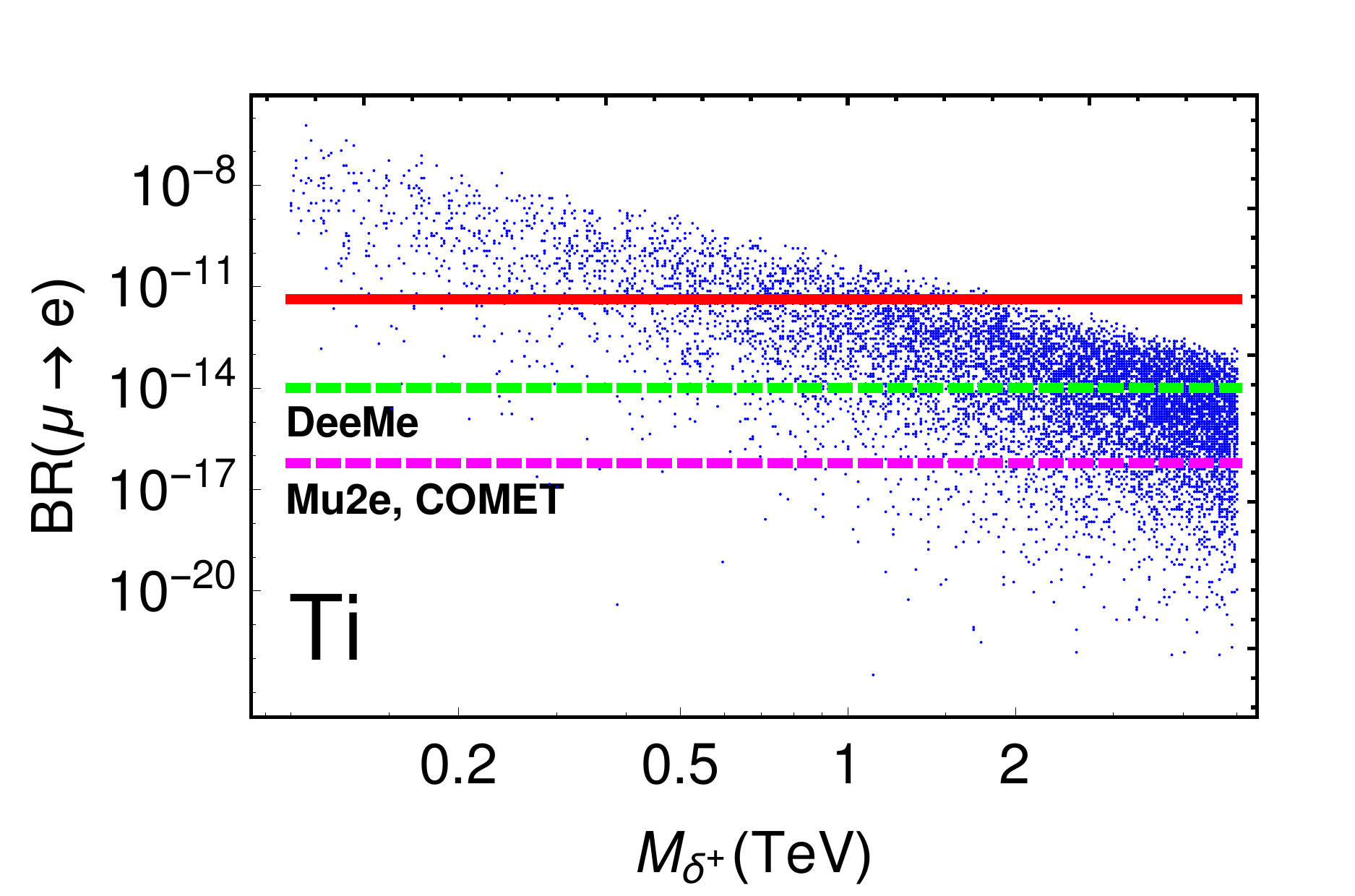}
 \includegraphics[width=0.45\linewidth,keepaspectratio=true,clip=true]{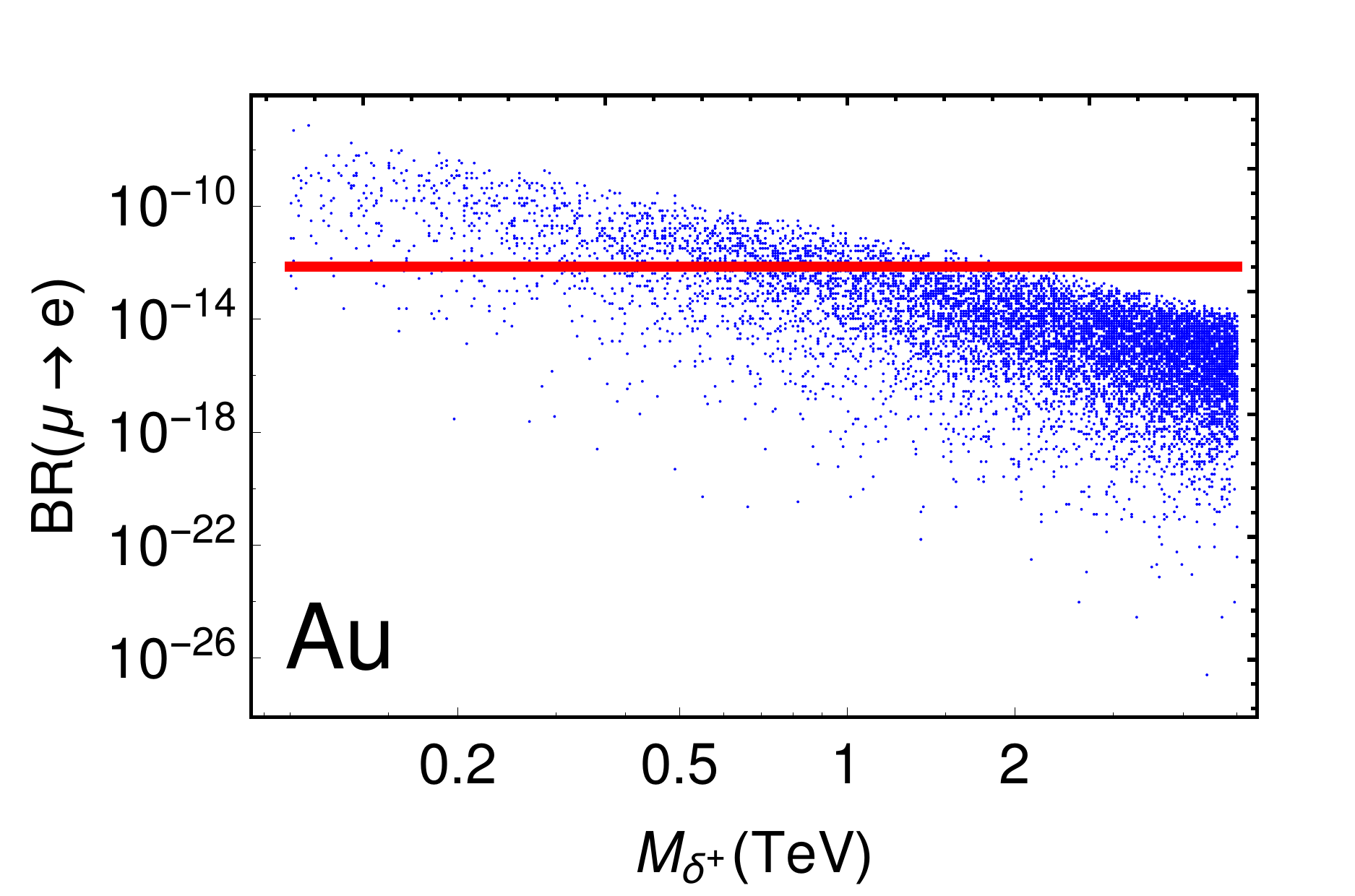}
 \includegraphics[width=0.45\linewidth,keepaspectratio=true,clip=true]{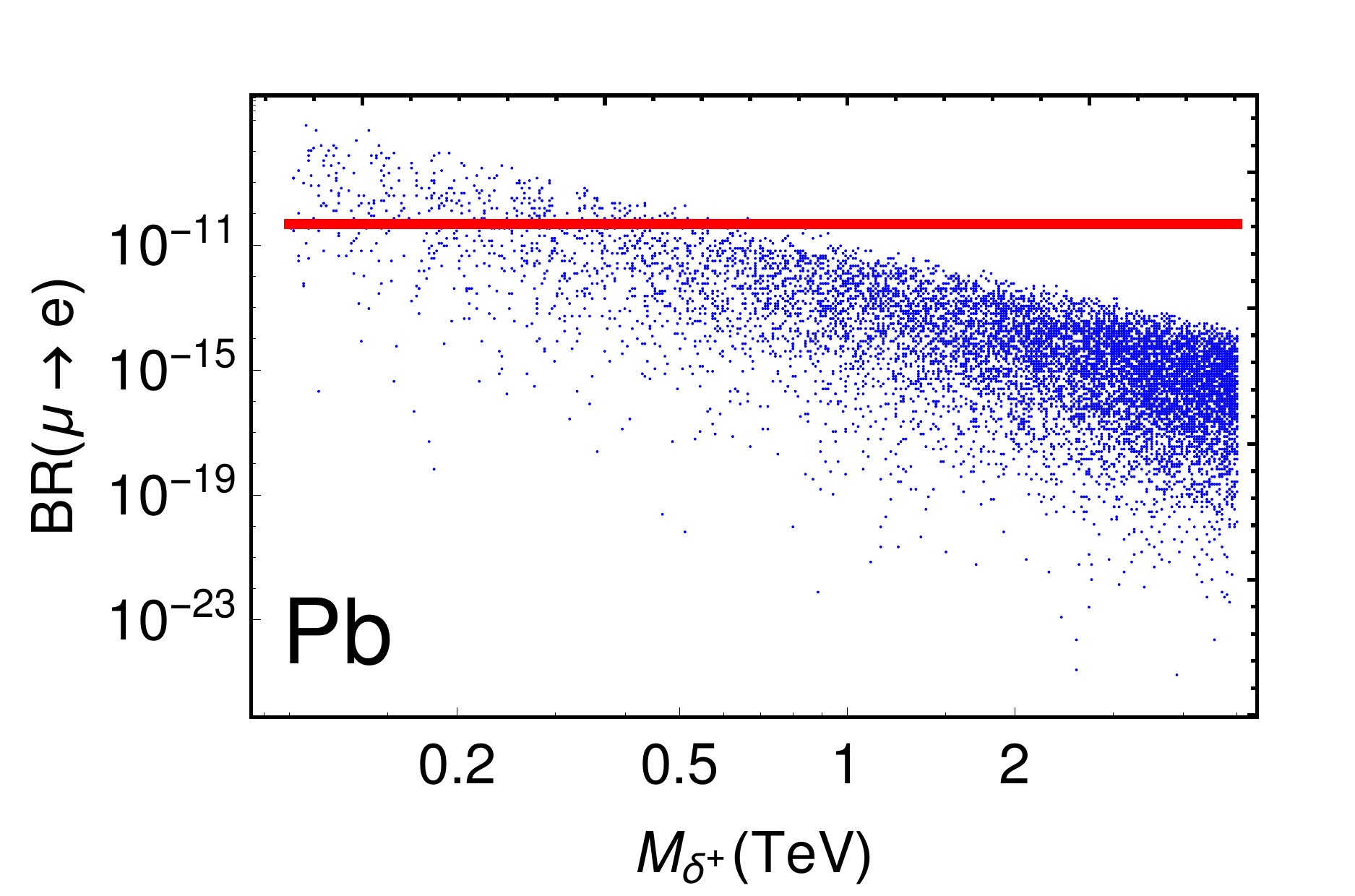}
  \includegraphics[width=0.45\linewidth,keepaspectratio=true,clip=true]{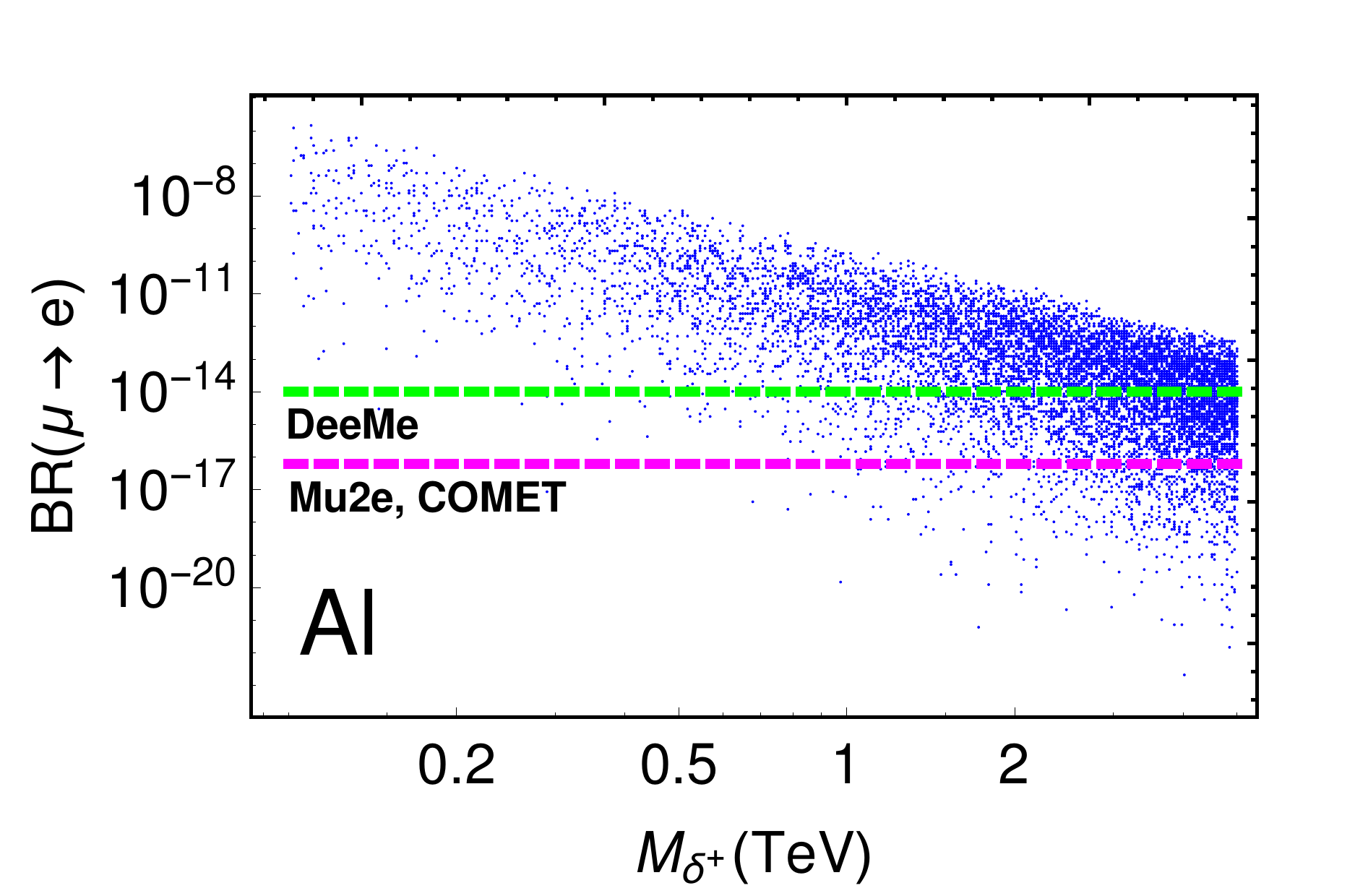}
 \caption{Prediction on the branching ratio of the processs $\mu-e$ conversion in different nuclei as a function of the mass of the 
 singly charged Higgs. Here, the Yukawa couplings $\lambda_R$ range from 0 to 0.1 and $\Delta$ from 0 to $\Delta_{\text{upper}}$ randomly.
 The red lines show the upper experimental bound for $\mu-e$ conversion, according to the nucleus. The dashed lines represent the projected sensitivities of the experiments DeeMe at J-PARC with $10^{-14}$~\cite{Aoki:2010zz} (green line), and COMET at J-PARC with $10^{-16}$~\cite{Cui:2009zz} and Mu2e at Fermilab with $6\times 10^{-17}$~\cite{Morescalchi:2016uks} (magenta line).}
 \label{mutoeisotops}
 \end{figure}
\begin{table}[h] 
  \vspace{0.5cm}
 \begin{tabular}{| c | c | c | c | c |}
  \hline
 Isotop & $\,\,D$~\cite{Kitano:2002mt}$\,\,$ & $\,\,V^{(p)}$~\cite{Kitano:2002mt}$\,\,$ & $\,\,V^{(n)}$~\cite{Kitano:2002mt}$\,\,$ & $\Gamma_{\text{capt}} (10^6\,s^{-1})$~\cite{Suzuki:1987jf} \\
  \hline
  $^{27}_{13}\text{Al}$  & 0.360 & 0.0160 & 0.0171 & 0.69 \\
  $^{48}_{22}\text{Ti}$  & 0.0867 & 0.0398 & 0.0482 & 2.59\\
  $^{197}_{79}\text{Au}$ & 0.178 & 0.0917  & 0.127 & 13.07 \\
  $^{207}_{82}\text{Pb}$ & 0.160 & 0.0828  & 0.119 & 13.45 \\
  \hline
  \end{tabular}
 \caption{For the dimensionless integrals, the average of the numerical values computed through different methodes has been taken (see Ref.~\cite{Kitano:2002mt}).}
 \label{Tab_mutoe}
\end {table}

Finally, the branching ratio is usually expressed as the conversion rate normalized by the muon capture rate
\begin{equation}
 \displaystyle B_{\mu \to e}(Z)=\frac{\Gamma_{\text{conv}}(Z,A)}{\Gamma_{\text{capt}}(Z,A)}.
\end{equation}
In Fig.~\ref{mutoeisotops} we show the prediction on the $\mu \to e$ process in the context of our left-right symmetric model as a function of the charged scalar 
mass for the isotopes $^{27}_{13}\text{Al}$, $^{48}_{22}\text{Ti}$, $^{197}_{79}\text{Au}$ and $^{207}_{82}\text{Pb}$. 
The current experimental bounds are represented by the red line and with dashed lines we show the projected limits. As one can see, the 
current bounds constrain only a small region of the parameter space, while the projected bounds will constrain the model in a significant way.
It is important to emphasize that in this theory one can have large lepton flavour violating effects in a consistent way in agreement 
with all experimental constrains. Then, combining the results for the $e_i \to e_j \gamma$ and the $\mu \to e$ conversion one can hope to test these predictions in the near future.

In this model one can have new contributions to neutrinoless double beta decay once the singly charged Higgs, $\delta^{\pm}$, 
mixes with the charged Higgses in the bi-doublet Higgs. Since the charged Higgses in the bi-doublet have to be heavy to avoid large 
flavour violation in the quark sector and the mixing angle is small the contribution to neutrinoless double beta decay is very small.
Therefore, here one has the usual contributions in the context of the left-right theory but taking into account that the right-handed neutrinos are light.
We have investigated the $e_i^{\pm} \to e^{\pm}_j e^{\mp}_k e^\pm_l$, and the predictions are quite below the current and projected experimental bounds. 
%
\section{Summary}
%
We have discussed the main features of a simple left-right symmetric theory~\cite{LRnew} where the neutrinos are Majorana fermions and their masses are generated at one-loop level.
This theory predicts that the right-handed neutrinos are generically light and the existence of new interactions violating lepton number which can give rise to large contributions to flavor violating processes such as $\mu \to e \gamma$ and $\mu \to e$ conversion.  We have investigated the lepton flavour violating contributions mediated by the new charged gauge bosons $W_R^\pm$. These contributions 
are very small in our model due to the GIM suppression. However, in other models where the right-handed neutrinos are heavy and close in mass to the $W_R^{\pm}$ 
gauge bosons, these contributions can be very large motivating the search for the LFV processes.

Our left-right symmetric theory predicts the existence of a new singly charged Higgs which can give rise to very large flavour number violation effects. Since the charged scalar could be in principle 
very light due to the lack of strong collider constrains, one can have large contributions to LFV processes such as $\mu \to e \gamma$ and $\mu \to e$ conversion. We have shown that both 
processes provide non-trivial bounds on this model and thanks to the possibility of improving the experimental bounds in the near future one can hope to test these predictions.
Together with the usual collider signatures from the $W_R^\pm$ and right-handed neutrinos decays, one can use the predictions for lepton flavour number violating processes to realize the testability of this theory.\\

{\it Aknowledgments}: The work of C.M. has been supported in part by the Spanish
Government and ERDF funds from the EU Commission [Grants
No. FPA2014-53631-C2-1-P and SEV-2014-0398] and ``La Caixa-Severo Ochoa" scholarship. 
%
\section{Appendices}
\subsection{Form factors}
%
The relevant coefficients for the study of $\mu \to e$ conversion are given by
\begin{eqnarray}
C_{VR}^{(q)}&& = \frac{Q_qe^2}{4\pi^2}\frac{1}{t}\sum_{N_i}\sum_{c,d}(\lambda_R^*)^{ce}\lambda_R^{d\mu}V_N^{ci}(V_N^*)^{di}
\left(\frac{3}{4}+m_{\delta^+}^2\left(\frac{1}{2(m_{\delta^+}^2-m_{N_i}^2)}-\frac{1}{t}\right)\right.\nonumber\\
&&+\frac{m_{N_i}^2}{t}+
\frac{\left((m_{N_i}^2-m_{\delta^+}^2)^2+m_{N_i}^2t\right)}{t} C_0[0,0,t,m_{\delta^+},m_{\delta^+},m_{N_i}] \nonumber \\
 &&  \left.+\text{Log}\left(\frac{m_{N_i}^2}{m_{\delta^+}^2}\right)\frac{2m_{N_i}^6+2m_{N_i}^2m_{\delta^+}^4+m_{N_i}^4(t-4m_{\delta^+}^2)}{2(m_{N_i}^2-m_{\delta^+}^2)^2t}+\frac{(2m_{N_i}^2-2m_{\delta^+}^2+t)\Lambda[t,m_{\delta^+},m_{\delta^+}]}{2t}\right), \nonumber \\
\end{eqnarray}
and
\begin{eqnarray}
C_{DR} &=& \frac{e}{8\pi^2}\sum_{N_i}\sum_{c,d}(\lambda_R^*)^{ce}\lambda_R^{d\mu}V_N^{ci}(V_N^*)^{di}\nonumber\\
&&\left(\frac{12m_{N_i}^4+12m_{\delta^+}^4-3m_{\delta^+}^2\,t+m_{N_i}^2(5t-24m_{\delta^+}^2)}{4(m_{N_i}^2-m_{\delta^+}^2)t^2} +\frac{(6m_{N_i}^2+6m_{\delta^+}^2+t)\,\Lambda[t,m_{\delta^+},m_{\delta^+}]}{2t^2}\right.\nonumber\\
&&+m_{N_i}^2\text{Log}\left(\frac{m_{N_i}^2}{m_{\delta^+}^2}\right)\frac{6m_{N_i}^4-6m_{\delta^+}^4-2m_{\delta^+}^2t+m_{N_i}^2(t-12m_{\delta^+}^2)}{2(m_{N_i}^2-m_{\delta^+}^2)^2 t^2}\nonumber\\
&&\left.+\frac{3m_{N_i}^4+3m_{\delta^+}^4-m_{\delta^+}^2t+m_{N_i}^2(2t-6m_{\delta^+}^2)}{t^2}C_0[0,0,t,m_{\delta^+},m_{\delta^+},m_{N_i}]\right),
\end{eqnarray}
where 
\begin{equation}
\Lambda[t,m_{\delta^+},m_{\delta^+}]\equiv \displaystyle \sqrt{1-\frac{4m_{\delta^+}^2}{t}}\text{Log}\left(\frac{2m_{\delta^+}^2}{2m_{\delta^+}^2-\left(1+\sqrt{1-\frac{2m_{\delta^+}^2}{t}}\right)t}\right),
\end{equation}
and the Passarino$-$Veltman $C_0$ function is given by
\begin{eqnarray}
 &&C_0[0,0,t,m_{\delta^+},m_{\delta^+},m_{N_i}]=\frac{1}{t}\left[-\text{DiLog}\left(\frac{(m_{N_i}^2-m_{\delta^+}^2)^2}{(m_{N_i}^2-m_{\delta^+}^2)^2+m_{N_i}^2t},t\right) \right. \nonumber \\ 
 &&+\text{DiLog}\left(\frac{(m_{N_i}^2-m_{\delta^+}^2)(m_{N_i}^2-m_{\delta^+}^2+t)}{(m_{N_i}^2-m_{\delta^+}^2)^2+m_{N_i}^2t},(m_{N_i}-m_{\delta^+})t(m_{N_i}^2-m_{\delta^+}^2+t)\right) \nonumber \\
 &&+\text{DiLog}\left(\frac{2(m_{N_i}-m_{\delta^+})(m_{N_i}+m_{\delta^+})}{2m_{N_i}^2-2m_{\delta^+}^2+t-\sqrt{t(t-4m_{\delta^+}^2)}},(m_{N_i}-m_{\delta^+})t\right) \nonumber \\
 &&-\text{DiLog}\left(-\frac{2(m_{N_i}^2-m_{\delta^+}^2+t)}{2m_{\delta^+}^2-2m_{N_i}^2-t+\sqrt{t(t-4m_{\delta^+}^2)}},(m_{N_i}^2-m_{\delta^+}^2+t)t\right) \nonumber \\
 &&+\text{DiLog}\left(\frac{2(m_{N_i}-m_{\delta^+})(m_{N_i}+m_{\delta^+})}{2m_{N_i}^2-2m_{\delta^+}^2+t+\sqrt{t(t-4m_{\delta^+}^2)}},(m_{\delta^+}-m_{N_i})t\right)\nonumber \\
  &&- \left. \text{DiLog}\left(\frac{2(m_{N_i}^2-m_{\delta^+}^2+t)}{2m_{N_i}^2-2m_{\delta^+}^2+t+\sqrt{t(t-4m_{\delta^+}^2)}}, t(m_{\delta^+}^2-m_{N_i}^2-t)\right)\right].
\end{eqnarray}
However, in the elastic limit $t\sim m_\mu^2$, one can neglect the variable t since $m_\mu^2 \sim 0$ compared to $m_{\delta^+}$ runing inside the loop. Hence, $C_{DR}$ corresponds to the amplitude 
of $\mu \to e \gamma$ given by Eq.(\ref{ARS}) with some arrangements of the factors to be consistent with the notation used in Eqs.(\ref{effective_lagrangian}) and (\ref{conversion_rate}),
\begin{equation}
 C_{DR} = \frac{e}{8\pi^2}\frac{1}{m_{\delta^+}^2}\sum_i\sum_{c,d}(\lambda_R^*)^{ce}\lambda_R^{d\mu}V_N^{ci}(V_N^*)^{di}\,G\left(
 \frac{m_{N_i}^2}{m_{\delta^+}^2}\right),
\end{equation}
where the function $G(x)$ is defined in Eq.(\ref{G}), and the vectorial form factor becomes zero in this limit, i.e.
$C_{VR}^{(q)}= 0$.
We have used the Package-X~\cite{Patel:2016fam} to perform the loop calculations. 
%
\subsection{Relevant Feynman rules}
Here we list some of the most relevant Feynman rules for our study 
\begin{itemize}
 \item $\displaystyle \bar{\nu}_i \ e_j\  W_R^+: i\frac{g_R}{\sqrt{2}}(B^{\dagger})^{ij}\gamma_{\mu}P_R$,
 \item $\displaystyle \bar{N}_i \ e_j \ W_R^+: i\frac{g_R}{\sqrt{2}}(V_N^{\dagger})^{ij}\gamma_{\mu}P_R$,
 \item $\displaystyle \bar{\nu}_i \ e_j \ W_L^+: i\frac{g_L}{\sqrt{2}}(V_\nu^{\dagger})^{ij}\gamma_{\mu}P_L$,
 \item $\displaystyle \bar{N}_i \ e_j \ W_L^+: i\frac{g_L}{\sqrt{2}}(A^{\dagger})^{ij}\gamma_{\mu}P_L$,
 \item $\displaystyle \bar{\nu}_i \ e_j \ \delta^+: 2i\sum_c (B^*)^{ci}\lambda_R^{cj}P_R+V_\nu^{ci}\lambda_L^{cj}P_L$,
 \item $\displaystyle \bar{N}_i \ e_j \ \delta^+: 2i\sum_c (V_N^*)^{ci}\lambda_R^{cj}P_R+A^{ci}\lambda_L^{cj}P_L$.
\end{itemize}
Here we have neglected the mixing between the $W_L$ and $W_R$ gauge bosons for simplicity.
%


\end{document}